**The Complex Rotational Light Curve of (385446) Manwë-Thorondor, a Multi-Component Eclipsing System in the Kuiper Belt**

short title: The Light Curve of (385446) Manwë-Thorondor

David L. Rabinowitz[1], Susan D. Benecchi[2], William M. Grundy[3], Anne J. Verbiscer[4], Audrey Thirouin[3]

[1]Yale University, Center for Astronomy and Astrophysics, P.O. Box 208120, New Haven, CT 06520-8120, United States
[2]Planetary Science Institute, 1700 E. Fort Lowell, Suite #106, Tucson, AZ 85719, United States
[3]Lowell Observatory, 1400 W. Mars Hill Road, Flagstaff, AZ 86001, United States
[4]University of Virginia, Department of Astronomy, PO Box 400325, Charlottesville, VA 22904, United States

## Abstract

Kuiper Belt Object (385446) Manwë-Thorondor is a multi-object system with mutual events predicted to occur from 2014 to 2019. To detect the events, we observed the system at 4 epochs (UT 2016 Aug 25 and 26, 2017 Jul 22 and 25, 2017 Nov 9, and 2018 Oct 6) in g, r, and VR bands using the 4-m SOAR and the 8.1-m Gemini South telescopes at Cerro Pachón, Chile and Lowell Observatory's 4.3-m Discovery Channel Telescope at Happy Jack, Arizona. These dates overlap the uncertainty range (±0.5 d) for four inferior events (Thorondor eclipsing Manwë). We clearly observe variability for the unresolved system with a double-peaked period 11.88190 ± 0.00005 h and ~0.5 mag amplitude together with much longer-term variability. Using a multi-component model, we simultaneously fit our observations and earlier photometry measured separately for Manwë and Thorondor with the Hubble Space Telescope. Our fit suggests Manwë is bi-lobed, close to the "barbell" shape expected for a strengthless body with density ~0.8 g/cm$^3$ in hydrostatic equilibrium. For Manwë, we thereby derive maximum width to length ratio ~0.30, surface area equivalent to a sphere of diameter 190 km, geometric albedo 0.06, mass 1.4x10$^{18}$ kg, and spin axis oriented ~75 deg from Earth's line of sight. Changes in Thorondor's brightness by ~0.6 mag with ~300-d period may account for the system's long-term variability. Mutual events with unexpectedly shallow depth and short duration may account for residuals to the fit. The system is complex, providing a challenging puzzle for future modeling efforts.

## 1. INTRODUCTION

The distant body (385446) Manwë-Thorondor is one of ~50 known Kuiper Belt Objects (KBOs) in a 7:4 mean motion resonance with Neptune[1]. The orbits of bodies within and near this resonance evolve over long time scales, leading to a mixing of the cold and hot (i.e. low and high inclination) classical populations (Lykawka & Mukai 2005, Volk & Malhotra 2011). Because these two populations are known to have distinct physical properties, with the cold population dominated by objects with very red optical colors (Peixhinho, Lacerda, & Jewitt 2008, Fornasier, Barucci, & de Bergh 2009, Peixhinho et al. 2015) and a large fraction of binary systems (Noll et al. 2008), assessing the physical properties of the KBOs within and near the 7:4 resonance constrains the extent to which the two populations have mixed. For example, Sheppard (2012) measured optical colors for 11 KBOs in the 7:4 resonance, including Manwë-Thorondor, and found all but one with very red optical colors similar to those observed for the cold population. However, other than their colors and orbits, little else is known about the physical composition of these resonant bodies.

Manwë-Thorondor is therefore of particular interest, not only because the mutual orbit of the bodies fixes the mass of the system, but also because at recent epochs Manwë and Thorondor have been alternately eclipsing and occulting each other. Using Hubble Space Telescope (HST) observations obtained from 2006 to 2013, Grundy et al. (2011, 2014) precisely determined the shape and orientation of the binary orbit and the system mass, $m_{sys}$ (Table 1 lists the orbit parameters). From these results they predicted mutual events with duration ~12 h occurring four to five times per year from 2014 to the end of 2019, with the timing of these events uncertain by ~1/2 day. Among the KBOs, the only well-separated eclipsing binaries previously observed have been Pluto-Charon (Buie, Tholen, & Horne 1992) and Sila-Nunam (Grundy et al. 2012, Verbiscer et al. 2013, Benecchi et al. 2014, Rabinowitz et al. 2014). Measuring the time dependence and depth of these events presents a unique opportunity to accurately determine the size, density, and albedo of the components, and to detect local variations in their surface color or albedo as one body is gradually hidden and then exposed by the other.

In this paper we report photometric observations of the Manwë-Thorondor system, taken with the dual goals of constraining the rotational properties and capturing at least one of the predicted mutual events. The HST observations previously reported by Grundy et al. (2012) show that Manwë and Thorondor differ in brightness by ~1.2 mag on average, with this difference changing significantly over time because of independent variability of 0.5 and 0.7 mag for Manwë and Thorondor respectively. Owing to the sparse and scattered epochs of the HST observations, Grundy et al. (2014) were unable to determine a rotational period for either body. They conclude, however, that it is unlikely the two bodies are tidally locked in a shared rotation state. From dynamical arguments they suggest Thorondor may be chaotically rotating or else spinning very rapidly. Given these considerations, we would expect our new photometric observations to reveal an irregular light curve for the system, with ~0.2-mag brightness fluctuations randomly phased with a larger-amplitude ( ~0.5 mag) periodic component.

[1] The Deep Ecliptic Survey Object Classifications compiled by M. Buie, Southwest Res. Inst. (https://www.boulder.swri.edu/~buie/kbo/desclass.html)

**Table 1**. Binary Orbital Parameters for Manwë-Thorondor[a]

| | |
|---|---|
| **orbital period** | 110.176 ± 0.018 d |
| **semimajor axis** | 6674 ± 41 km |
| **eccentricity** | 0.5632 ± 0.0070 |
| **$m_{sys}$** | 1.941 ± 0.036 x $10^{18}$ kg |
| **inclination** | 25.58 ± 0.23 deg |
| **ascending node** | 163.56 ± 0.78 deg |

Table Notes: [a]Orbital elements are for Thorondor relative to Manwë (Grundy et al. 2014) and referred to the J2000 equatorial frame

Section 2 of this paper presents our observations and reduction procedures. In Section 3, we analyze the observations using two different methods. The first approach is to fit a sinusoidal function to the dominant periodic component of the light curve. Motivated by the limitations of this approach, we then use a more complex, multi-component model to predict and fit the observations, with the assumption that Manwë is a bi-lobed object with the barbell shape expected for a contact binary. We discuss the implications of both model fits in Section 4 and 5.

## 2. OBSERVATIONS AND REDUCTION PROCEDURES

The observations we report group into four epochs. Each epoch includes at least one night targeting a predicted inferior event (Thorondor and/or its shadow moving in front of or over Manwë). Superior events were not targeted because of their coincidence with bright moon nights. When possible, observations were also made on a second (ideally consecutive) night when no mutual event was expected. With the SOAR and Gemini telescopes, observations were made alternating Sloan g and r filters in order to measure the dependence of color on rotational phase. Table 2 lists the observing circumstances for each epoch along with the expected mid-times and maximum depths of the targeted mutual event, and the target's solar phase angle (θ) and heliocentric and geocentric distances ($r_{hel}$ and $r_{geo}$). Given the large uncertainty in the predicted event times, there is a significant chance for the events to occur in daytime before or after the targeted evening.

**Table 2. Observing Circumstances**

| Epoch No. | Date | Telescope | Filters | Obs Start | Obs End | Event Midpt | Event Depth (mag) | θ (deg) | $r_{hel}$ (AU) | $r_{geo}$ (AU) |
|---|---|---|---|---|---|---|---|---|---|---|
| 1 | 2016 Aug 25 | SOAR | g,r | 2:40 | 9:36 | --- | --- | 0.48 | 43.68 | 42.73 |
| | 2016 Aug 26 | SOAR | g,r | 2:10 | 9:07 | 5:15 | 0.66 | 0.46 | 43.68 | 42.73 |
| 2 | 2017 Jul 22 | Gemini | g,r | 5:16 | 10:33 | 20:01 | 0.34 | 1.10 | 43.58 | 43.00 |
| | 2017 Jul 25 | Gemini | g,r | 5:45 | 8:24 | --- | --- | 1.06 | 43.58 | 43.00 |
| 3 | 2017 Nov 9 | DCT | r | 2:09 | 5:02 | 23:36 | 0.62 | 1.05 | 43.55 | 42.95 |
| 4 | 2018 Oct 6 | DCT | VR | 3:25 | 9:15 | 14:56 | 0.05 | 0.40 | 43.45 | 42.50 |

Notes: All date and times are UT. Event midpoints and depths are predicted values assuming spherical shaped bodies with Lambertian light-scattering properties and nominal diameters 160 and 92 km for Manwë and Thorondor, respectively. Event parameters are unlisted if no event was expected within 12 hours of the observation start or end time. Event midpoints are uncertain by ~0.5 day, and event durations are ~0.5 day. All events listed above are inferior events with Thorondor in front of Manwë. See Grundy et al. (2014) http://www2.lowell.edu/users/grundy/tnbs/385446_2003_QW111_Manwë-Thorondor_mutual_events.html.

## 2.1 Epoch 1 – SOAR Telescope Observations

The epoch 1 observations reported here were obtained in the SDSS r and g bands on two consecutive nights (2016 Aug 25 and 26 UT) using the SOAR Optical Imager (SOI) on the 4-m SOAR telescope at Cerro Pachón, Chile. The imager consists of a mini-mosaic with two thinned, back illuminated E2V 2kx4k charge-coupled devices (CCDs), together covering a 5.26 arcmin square field. We used the 2x2-binned readout mode, yielding a pixel scale of 0.154 arcsec. On both nights we took advantage of the remote-observing capabilities of the SOAR telescope, controlling the camera and telescope from our home institutions in the USA. Operators and staff at the observatory provided instrument support and monitored the telescope, camera, and weather status throughout the night.

On both nights we observed Manwë-Thorondor at 8- to 10-min intervals over a period of 8 hours (02:00 to 10:00 h UT), with a nominal observing cadence of four r-band exposures followed by one in g band. Most exposures were fixed at 450 and 600 s for the r and g bands, respectively. Except for small dithers between exposures, we targeted the same field on both nights. This allowed us to use an identical set of field stars to calibrate all the photometry. The two-night motion of Manwë-Thorondor was 1.5 arcmin, well within the field covered by the camera. We also oriented the camera so that the target did not move across the gap between the two CCDs in the array. Both nights were clear with seeing variable from 0.5 to 2.0 arcsec (median 1.2 arcsec).

Our second night was nearly optimally timed to cover the mutual event with an expected mid-time, $t_{mid}$ = 05:25 h UT Aug 26, and duration ~10 h. Here we define mutual event duration to be the interval of time when the expected brightness diminution would exceed 0.1 mag (see Grundy et al. 2014). Given the 12-h uncertainty in the timing and our ~8 h observation windows, our likelihood of detecting at least half the event was ~29% (5.3% and 24.1% on the nights of Aug 25 and 26, respectively). For detecting at least a 10% diminution, our expected likelihood was ~64% (13% and 51% on the respective nights).

### 2.2 Epoch 2 – Gemini South Telescope Observations

The epoch 2 observations were obtained in SDSS r- and g-bands on 2017 Jul 22 and Jul 25 UT using the GMOS-S imager on the 8.1-m Gemini South telescope at Cerro Pachón (the second night was delayed by 2 nights due to bad weather). The time was awarded under Gemini's "Fast Turnaround Time" program. The imager is an array of three Hamumatsu 2048x4176 pixel CCDs with a field of 6144 x 4224 pixels covering 8.35 x 5.63 arcmin. With 2x2 binning, the pixel scale is 0.16 arcsec. The queued observations were obtained by onsite operators on both nights, and sequenced continuously while conditions were acceptable (most of night 1, part of night 2). We monitored the operations remotely with a Skype connection.

Sequences consisted of five r-band exposures (240 s each) followed by two g-band exposures (150 s each), repeated throughout the night. All exposures were dithered, with the same pointing used for all exposures on both nights. The target was centered in the field on the first night, but appeared near a gap between two detectors on the second night. Because of the dithering on the 2$^{nd}$ night, the target sometimes fell into the gap between two of the detectors. Both nights were marginally clear with seeing variable from 1.0 to 2.0 arcsec and with median values 1.2 arcsec (night 1) and 1.6 arcsec (night 2).

### 2.3 Epochs 3 and 4 – Lowell Observatory's Discovery Channel Telescope Observations

The epoch 3 observations were obtained on 2017 Nov 9 UT in r band using the Large Monolithic Imager (LMI) on Lowell Observatory's Discovery Channel Telescope (DCT) at Happy Jack AZ. The LMI is a single 6144 x 6160 pixel E2V CCD with 0.24 arcsec pixel scale (3x3 binned) covering a field 12.3 x 12.3 arcmin. Observations had been planned for the latter halves of both 2017 Nov 8 and 9 UT. Because of bad weather, only sparse observations were obtained, and only on the second night. All exposures were 600s. Seeing was 1 to 1.5 arcsec.

The epoch 4 observations were obtained 2018 Oct 6 UT in a wide VR band, again using the LMI on the DCT. Seeing was stable, between 1.2 and 1.5 arcsec. The target was too faint to observe using narrower-band filters.

### 2.4 SOAR and Gemini Reduction Procedures

To obtain the SOAR and Gemini light curves, we first used standard techniques to remove bias and flat field artifacts from the raw images, relying on bias exposures and twilight flats taken on both nights. For both the SOAR and Gemini data sets, each consisting of 2 nights of observations, we then used two different procedures to calibrate the data. With procedure 1 we used IRAF programs to obtain a relative calibration of the target magnitudes, with careful attention to account not only for the relative extinction and seeing from one exposure to the next, but also for the influence of faint field stars near the path of the target. With procedure 2 we obtained absolutely calibrated magnitudes, including color corrections, making use of a photometry pipeline that measures and calibrates source magnitudes for fields that are within the footprint of the Sloan Digital Sky Survey (SDSS). This second procedure is described in Rabinowitz et al. 2014.

We describe our implementation of procedure 1 in detail below.

Step 1: We determined the magnitude zero points for all the images on both nights relative to a chosen best image (least extinction, best seeing) by measuring the instrumental magnitudes for a fixed set of field stars common to all the images. We used a large aperture with diameter nearly double the worst seeing (2.5 arcsec for SOAR, 3.2 arcsec for Gemini) for the field star photometry, and also a smaller aperture comparable to the median seeing (1.2 arcsec for SOAR, 1.3 arcsec for Gemini) to determine an aperture correction for the target photometry (see step 4).

Step 2: For each night we constructed a median reference image by selecting the best images (least extinction, best seeing, deepest exposures), sky-subtracting them, shifting their pixels to achieve a common alignment, scaling them to compensate for their relative zero points, and then computing the median average at each pixel location.

Step 3: For each image acquired on a given night, we removed the field stars by subtracting the alternate night's median reference image after shifting each image to align with the reference, and scaling the reference to have the same zero point as the image. An alternate night's reference image was chosen to make sure there was no residual signal in the reference image at the location of the moving target.

Step 4: Using the same small aperture used in Step 1 to measure the field stars, we measured the magnitudes for the target in the reference-subtracted images. We then used relative zero points and aperture corrections determined from the field stars (step 1) to normalize the photometry.

For the r-band, we used procedure 1 to obtain a precise measure of the light curve shape and then relied on procedure 2 only to have an independent check on these measurements and to establish the magnitude offset and color terms required to transform the photometry into the Sloan system. For the g-band, because we had only one night of good data (Gemini) or because we did not have enough observations to construct a good median sky image for both nights (SOAR), we relied exclusively on the second procedure for all photometry.

For both procedures 1 and 2 we used a small aperture comparable to the median seeing to measure the target flux. Note that the motion of the target in the SOAR r-band images (nearly all with 450-sec exposure times) was 0.34 arcsec, leading to a negligible influence on the photometry (<1% loss in signal) given the much larger photometric aperture (1.5"). The g-band exposures were longer (600 sec), leading to a possible small trailing loss of ~1.5%. There was insignificant trailing loss in the short Gemini exposures.

For both the r and g photometry, we determined color corrections iteratively in two passes under the assumption that the corrections were the same for all images in a given band. In the first pass we made an initial estimate of g-r without color corrections, knowing this would require a later correction. In the second pass we used our estimate for g-r to determine the color corrections individually for each measurement. Then for each band we evaluated the mean color correction and used this to correct all the measurement from both nights. Our final estimate for g-r is derived from these color-corrected values for g and r.

2.5 DCT Reduction Procedures

The DCT observations were reduced using the standard procedures described by Thirouin & Sheppard (2019).

The r-band data collected on 2018 Nov 8 were taken through clouds in poor conditions. We include the few observations that could be calibrated to standards, but give them no weight in the foregoing analysis.

The VR-band data were calibrated to r-band standards. In the analysis below, we attempt to simultaneously fit the resulting r-band light curve along with our r-band photometry from previous epochs to a common model, ignoring any possible changes in color over time that might affect the relative calibration of the r and VR bands to the same magnitude system.

## 3. RESULTS AND ANALYSIS

3.1 Photometric Measurements

Table 3 lists all our magnitude measurements, M, along with Julian Date (JD) at the start of each exposure, the measurement error, the exposure time of the associated image, and the telescope. Light travel-time corrections relative to a common epoch (JD 2454300.0) were computed for each observation date and used in the analysis described below. For the r-band, the listed measurement errors do not include the systematic uncertainty from the absolute zero point calibration and color correction. They do include the uncertainties in the instrumental magnitudes and their relative normalization determined from the statistical noise in the measured sky and target flux. The zero point uncertainties for both the g- and r-band magnitudes are ~2%, based on unsaturated stars in the field with SDSS magnitude uncertainties < 10%. Measurements contaminated by cosmic ray hits or taken in poor observing conditions (wind bounce, high airmass, poor focus, or insufficient exposure time are excluded from the tables and the analysis described below.

**Table 3.** Photometric Observations of Manwë-Thorondor

| Telescope | JD-2450000 (d) | Filter | M (mag) | M err (mag) | Exp Time (s) | Fit Residuals (mag) |
|---|---|---|---|---|---|---|
| SOAR | 7625.60476 | r | 22.95 | 0.06 | 360 | -0.17 |
| SOAR | 7625.61123 | r | 23.31 | 0.07 | 600 | 0.17 |
| SOAR | 7625.62262 | r | 23.25 | 0.07 | 300 | 0.04 |
| SOAR | 7625.63440 | r | 23.39 | 0.07 | 300 | 0.09 |
| SOAR | 7625.64710 | r | 23.57 | 0.06 | 300 | 0.16 |
| SOAR | 7625.65456 | r | 23.42 | 0.06 | 300 | -0.05 |
| SOAR | 7625.65977 | r | 23.30 | 0.06 | 300 | -0.21 |
| SOAR | 7625.67192 | r | 23.54 | 0.05 | 300 | 0.01 |
| SOAR | 7625.67788 | r | 23.58 | 0.05 | 300 | 0.07 |
| SOAR | 7625.68234 | g | 24.18 | 0.10 | 300 | -0.12 |
| SOAR | 7625.68663 | g | 24.34 | 0.08 | 600 | 0.08 |
| SOAR | 7625.69516 | r | 23.52 | 0.04 | 600 | 0.15 |
| SOAR | 7625.70381 | r | 23.39 | 0.03 | 600 | 0.10 |
| SOAR | 7625.71162 | r | 23.22 | 0.03 | 600 | -0.01 |
| SOAR | 7625.71916 | r | 23.24 | 0.06 | 450 | 0.06 |
| SOAR | 7625.72603 | r | 23.14 | 0.04 | 450 | 0.00 |
| SOAR | 7625.73388 | r | 23.13 | 0.04 | 450 | 0.02 |
| SOAR | 7625.74064 | g | 24.08 | 0.11 | 450 | 0.18 |
| SOAR | 7625.74630 | r | 23.15 | 0.04 | 450 | 0.09 |
| SOAR | 7625.75235 | r | 23.04 | 0.04 | 450 | -0.01 |
| SOAR | 7625.75826 | r | 23.14 | 0.04 | 450 | 0.10 |
| SOAR | 7625.76394 | g | 23.88 | 0.09 | 600 | 0.03 |
| SOAR | 7625.77170 | r | 22.97 | 0.03 | 600 | -0.05 |
| SOAR | 7625.77955 | r | 22.96 | 0.04 | 450 | -0.06 |
| SOAR | 7625.78571 | r | 23.06 | 0.04 | 450 | 0.04 |
| SOAR | 7625.79210 | g | 23.75 | 0.10 | 600 | -0.08 |
| SOAR | 7625.79970 | r | 22.86 | 0.04 | 450 | -0.16 |
| SOAR | 7625.80653 | r | 22.93 | 0.04 | 450 | -0.09 |
| SOAR | 7625.81317 | r | 22.95 | 0.05 | 450 | -0.07 |
| SOAR | 7625.82271 | g | 23.76 | 0.09 | 600 | -0.09 |
| SOAR | 7625.83168 | g | 23.98 | 0.13 | 450 | 0.12 |
| SOAR | 7625.83990 | r | 23.05 | 0.04 | 450 | -0.02 |
| SOAR | 7625.84560 | r | 23.08 | 0.04 | 450 | -0.01 |
| SOAR | 7625.85132 | r | 23.21 | 0.05 | 450 | 0.11 |
| SOAR | 7625.85706 | g | 23.84 | 0.10 | 600 | -0.11 |
| SOAR | 7625.86465 | r | 23.22 | 0.05 | 450 | 0.04 |
| SOAR | 7625.87045 | r | 23.29 | 0.05 | 450 | 0.07 |
| SOAR | 7625.87884 | r | 23.32 | 0.05 | 450 | 0.04 |
| SOAR | 7625.88458 | r | 23.21 | 0.05 | 450 | -0.12 |
| SOAR | 7625.90024 | r | 23.52 | 0.07 | 450 | 0.06 |
| SOAR | 7626.59043 | r | 23.06 | 0.05 | 450 | -0.03 |
| SOAR | 7626.59870 | r | 23.12 | 0.05 | 450 | -0.01 |
| SOAR | 7626.60505 | r | 23.26 | 0.06 | 450 | 0.10 |
| SOAR | 7626.61116 | g | 24.24 | 0.10 | 600 | 0.22 |
| SOAR | 7626.62057 | r | 23.42 | 0.07 | 450 | 0.15 |
| SOAR | 7626.62595 | r | 23.48 | 0.05 | 450 | 0.17 |
| SOAR | 7626.63130 | r | 23.38 | 0.04 | 450 | 0.02 |
| SOAR | 7626.63669 | r | 23.39 | 0.06 | 450 | -0.02 |
| SOAR | 7626.65234 | r | 23.44 | 0.05 | 450 | -0.09 |
| SOAR | 7626.65772 | r | 23.52 | 0.05 | 450 | -0.02 |
| SOAR | 7626.66310 | r | 23.43 | 0.04 | 450 | -0.10 |
| SOAR | 7626.66846 | r | 23.61 | 0.05 | 450 | 0.10 |
| SOAR | 7626.67597 | g | 24.29 | 0.08 | 600 | 0.03 |
| SOAR | 7626.68593 | r | 23.28 | 0.06 | 450 | -0.08 |
| SOAR | 7626.69127 | r | 23.33 | 0.04 | 450 | 0.01 |
| SOAR | 7626.69668 | r | 23.23 | 0.04 | 450 | -0.04 |
| SOAR | 7626.70207 | r | 23.19 | 0.04 | 450 | -0.03 |
| SOAR | 7626.70784 | g | 24.03 | 0.06 | 600 | 0.02 |
| SOAR | 7626.71699 | r | 23.14 | 0.03 | 450 | 0.00 |
| SOAR | 7626.72234 | r | 23.16 | 0.03 | 450 | 0.04 |
| SOAR | 7626.72772 | r | 23.05 | 0.04 | 450 | -0.04 |
| SOAR | 7626.73314 | r | 23.10 | 0.03 | 450 | 0.03 |
| SOAR | 7626.73904 | g | 23.78 | 0.05 | 600 | -0.09 |
| SOAR | 7626.74748 | r | 23.00 | 0.04 | 450 | -0.04 |
| SOAR | 7626.75286 | r | 23.01 | 0.04 | 450 | -0.02 |
| SOAR | 7626.75822 | r | 22.99 | 0.03 | 450 | -0.03 |
| SOAR | 7626.76357 | r | 22.90 | 0.03 | 450 | -0.12 |
| SOAR | 7626.76980 | g | 23.79 | 0.06 | 600 | -0.05 |
| SOAR | 7626.77904 | r | 23.00 | 0.04 | 450 | -0.02 |
| SOAR | 7626.78442 | r | 23.05 | 0.04 | 450 | 0.03 |
| SOAR | 7626.78980 | r | 22.82 | 0.05 | 450 | -0.20 |
| SOAR | 7626.79515 | r | 22.96 | 0.04 | 450 | -0.06 |
| SOAR | 7626.80199 | g | 23.84 | 0.09 | 600 | 0.00 |
| SOAR | 7626.81075 | r | 22.98 | 0.05 | 450 | -0.05 |
| SOAR | 7626.81616 | r | 22.90 | 0.04 | 450 | -0.14 |
| SOAR | 7626.82157 | r | 23.06 | 0.04 | 450 | 0.01 |
| SOAR | 7626.82697 | r | 22.99 | 0.05 | 450 | -0.07 |
| SOAR | 7626.83286 | g | 23.83 | 0.09 | 600 | -0.07 |
| SOAR | 7626.84105 | r | 23.29 | 0.05 | 450 | 0.19 |
| SOAR | 7626.84645 | r | 23.25 | 0.06 | 450 | 0.12 |
| SOAR | 7626.85183 | r | 23.16 | 0.06 | 450 | 0.01 |
| SOAR | 7626.85721 | r | 23.17 | 0.05 | 450 | -0.02 |
| SOAR | 7626.86308 | g | 24.06 | 0.12 | 600 | 0.01 |
| SOAR | 7626.87463 | r | 23.48 | 0.06 | 450 | 0.16 |
| SOAR | 7626.88006 | r | 23.35 | 0.07 | 450 | -0.02 |
| SOAR | 7626.88544 | r | 23.53 | 0.08 | 450 | 0.11 |
| SOAR | 7626.89085 | r | 23.37 | 0.09 | 450 | -0.09 |
| Gemini | 7956.71597 | r | 23.13 | 0.04 | 240 | 0.08 |
| Gemini | 7956.71917 | r | 23.06 | 0.03 | 240 | 0.01 |
| Gemini | 7956.72237 | r | 22.96 | 0.05 | 240 | -0.08 |
| Gemini | 7956.72560 | r | 23.14 | 0.05 | 240 | 0.11 |
| Gemini | 7956.72884 | r | 22.99 | 0.05 | 240 | -0.04 |
| Gemini | 7956.73211 | g | 23.74 | 0.06 | 150 | -0.11 |
| Gemini | 7956.73427 | g | 23.83 | 0.07 | 150 | -0.02 |
| Gemini | 7956.73649 | r | 23.11 | 0.04 | 240 | 0.08 |
| Gemini | 7956.73970 | r | 23.03 | 0.03 | 240 | 0.00 |
| Gemini | 7956.74292 | r | 23.02 | 0.03 | 240 | 0.00 |
| Gemini | 7956.74616 | r | 23.08 | 0.04 | 240 | 0.06 |
| Gemini | 7956.74935 | r | 22.95 | 0.04 | 240 | -0.08 |
| Gemini | 7956.75262 | g | 23.84 | 0.09 | 150 | -0.01 |
| Gemini | 7956.75484 | g | 23.89 | 0.06 | 150 | 0.05 |
| Gemini | 7956.75706 | r | 23.02 | 0.03 | 240 | -0.01 |
| Gemini | 7956.76030 | r | 23.07 | 0.03 | 240 | 0.05 |
| Gemini | 7956.76352 | r | 23.07 | 0.03 | 240 | 0.05 |
| Gemini | 7956.76677 | r | 23.09 | 0.04 | 240 | 0.07 |
| Gemini | 7956.77004 | r | 23.17 | 0.04 | 240 | 0.14 |
| Gemini | 7956.77545 | g | 24.01 | 0.06 | 150 | 0.16 |
| Gemini | 7956.77766 | r | 23.00 | 0.04 | 240 | -0.03 |
| Gemini | 7956.78090 | r | 23.14 | 0.05 | 240 | 0.11 |
| Gemini | 7956.78413 | r | 23.06 | 0.03 | 240 | 0.02 |
| Gemini | 7956.78736 | r | 23.04 | 0.02 | 240 | 0.00 |
| Gemini | 7956.79057 | r | 23.03 | 0.04 | 240 | -0.01 |
| Gemini | 7956.79383 | g | 23.74 | 0.06 | 150 | -0.12 |
| Gemini | 7956.79602 | g | 23.87 | 0.06 | 150 | 0.00 |
| Gemini | 7956.79823 | r | 23.00 | 0.03 | 240 | -0.06 |
| Gemini | 7956.80144 | r | 23.11 | 0.03 | 240 | 0.04 |
| Gemini | 7956.80464 | r | 23.08 | 0.04 | 240 | 0.00 |
| Gemini | 7956.80786 | r | 23.11 | 0.10 | 240 | 0.02 |
| Gemini | 7956.81109 | r | 23.06 | 0.03 | 240 | -0.04 |
| Gemini | 7956.81435 | g | 23.97 | 0.11 | 150 | 0.04 |
| Gemini | 7956.81654 | g | 24.13 | 0.07 | 150 | 0.19 |
| Gemini | 7956.81875 | r | 23.19 | 0.04 | 240 | 0.05 |
| Gemini | 7956.82197 | r | 23.13 | 0.04 | 240 | -0.03 |
| Gemini | 7956.82517 | r | 23.20 | 0.04 | 240 | 0.03 |
| Gemini | 7956.82843 | r | 23.35 | 0.05 | 240 | 0.16 |
| Gemini | 7956.83170 | r | 23.27 | 0.03 | 240 | 0.06 |
| Gemini | 7956.83498 | g | 24.12 | 0.08 | 150 | 0.06 |
| Gemini | 7956.83716 | g | 24.03 | 0.06 | 150 | -0.04 |
| Gemini | 7956.83938 | r | 23.38 | 0.06 | 240 | 0.11 |
| Gemini | 7956.84263 | r | 23.30 | 0.05 | 240 | 0.01 |
| Gemini | 7956.84590 | r | 23.31 | 0.04 | 240 | -0.01 |
| Gemini | 7956.84912 | r | 23.41 | 0.05 | 240 | 0.06 |
| Gemini | 7956.85234 | r | 23.55 | 0.06 | 240 | 0.17 |
| Gemini | 7956.85560 | g | 24.39 | 0.10 | 150 | 0.16 |
| Gemini | 7956.85780 | g | 24.46 | 0.15 | 150 | 0.21 |
| Gemini | 7956.86113 | r | 23.50 | 0.07 | 240 | 0.04 |
| Gemini | 7956.86492 | r | 23.52 | 0.06 | 240 | 0.03 |
| Gemini | 7956.86816 | r | 23.51 | 0.05 | 240 | -0.01 |
| Gemini | 7956.87138 | r | 23.52 | 0.05 | 240 | -0.02 |
| Gemini | 7956.87460 | r | 23.62 | 0.07 | 240 | 0.07 |
| Gemini | 7956.87786 | g | 24.58 | 0.13 | 150 | 0.19 |
| Gemini | 7956.88006 | g | 24.31 | 0.11 | 150 | -0.07 |
| Gemini | 7956.88227 | r | 23.42 | 0.04 | 240 | -0.14 |
| Gemini | 7956.88548 | r | 23.42 | 0.05 | 240 | -0.13 |
| Gemini | 7956.88868 | r | 23.56 | 0.05 | 240 | 0.03 |
| Gemini | 7956.89192 | r | 23.42 | 0.04 | 240 | -0.08 |
| Gemini | 7956.89519 | r | 23.53 | 0.05 | 240 | 0.06 |
| Gemini | 7956.89845 | g | 24.13 | 0.08 | 150 | -0.13 |
| Gemini | 7956.90061 | g | 24.10 | 0.09 | 150 | -0.14 |
| Gemini | 7956.90282 | r | 23.34 | 0.03 | 240 | -0.06 |
| Gemini | 7956.90605 | r | 23.36 | 0.06 | 240 | -0.01 |
| Gemini | 7956.90928 | r | 23.35 | 0.04 | 240 | 0.00 |
| Gemini | 7956.91249 | r | 23.39 | 0.05 | 240 | 0.08 |
| Gemini | 7956.91568 | r | 23.41 | 0.04 | 240 | 0.13 |
| Gemini | 7956.92113 | g | 24.16 | 0.08 | 150 | 0.09 |
| Gemini | 7956.92335 | r | 23.34 | 0.03 | 240 | 0.11 |
| Gemini | 7956.92657 | r | 23.29 | 0.05 | 240 | 0.09 |
| Gemini | 7956.92979 | r | 23.24 | 0.03 | 240 | 0.06 |
| Gemini | 7956.93299 | r | 23.23 | 0.05 | 240 | 0.06 |
| Gemini | 7956.93620 | r | 23.17 | 0.04 | 240 | 0.02 |
| Gemini | 7956.94164 | g | 24.16 | 0.25 | 150 | 0.22 |
| Gemini | 7959.74151 | r | 22.83 | 0.04 | 240 | -0.15 |
| Gemini | 7959.74803 | g | 23.87 | 0.11 | 150 | 0.02 |
| Gemini | 7959.75019 | g | 23.68 | 0.08 | 150 | -0.17 |
| Gemini | 7959.75240 | r | 23.10 | 0.04 | 240 | 0.10 |
| Gemini | 7959.75561 | r | 22.97 | 0.04 | 240 | -0.17 |
| Gemini | 7959.75883 | r | 23.06 | 0.03 | 240 | -0.03 |
| Gemini | 7959.77072 | g | 23.84 | 0.09 | 150 | -0.05 |
| Gemini | 7959.77618 | r | 23.17 | 0.03 | 240 | 0.03 |
| Gemini | 7959.77940 | r | 23.16 | 0.03 | 240 | -0.01 |
| Gemini | 7959.78591 | r | 23.43 | 0.05 | 240 | -0.05 |
| Gemini | 7959.78918 | g | 24.01 | 0.10 | 150 | 0.05 |
| Gemini | 7959.79358 | r | 23.17 | 0.04 | 240 | 0.01 |
| Gemini | 7959.79681 | r | 23.54 | 0.06 | 240 | -0.01 |
| Gemini | 7959.80001 | r | 23.19 | 0.04 | 240 | -0.06 |
| Gemini | 7959.80646 | r | 23.20 | 0.04 | 240 | -0.18 |
| Gemini | 7959.81193 | g | 24.14 | 0.08 | 150 | 0.04 |
| Gemini | 7959.81736 | r | 23.34 | 0.05 | 240 | -0.08 |
| Gemini | 7959.82056 | r | 21.75 | 0.21 | 240 | -0.09 |
| Gemini | 7959.82379 | r | 23.55 | 0.10 | 240 | -0.05 |
| Gemini | 7959.82701 | r | 23.29 | 0.05 | 240 | 0.31 |
| Gemini | 7959.83029 | g | 24.41 | 0.11 | 150 | 0.14 |
| Gemini | 7959.84433 | r | 23.89 | 0.07 | 240 | -0.08 |
| Gemini | 7959.85089 | g | 24.38 | 0.14 | 150 | -0.01 |
| Gemini | 7959.85306 | g | 24.27 | 0.16 | 150 | -0.11 |
| Gemini | 7959.85527 | r | 23.80 | 0.07 | 240 | -0.08 |
| DCT | 8066.58801 | r | 23.03 | 0.12 | 600 | -0.28 |
| DCT | 8066.60257 | r | 23.10 | 0.11 | 600 | -0.12 |
| DCT | 8066.60970 | r | 23.02 | 0.08 | 600 | -0.17 |
| DCT | 8066.65462 | r | 23.34 | 0.13 | 600 | 0.21 |
| DCT | 8066.68797 | r | 23.15 | 0.09 | 600 | 0.01 |
| DCT | 8397.64232 | VR | 23.08 | 0.06 | 600 | -0.05 |
| DCT | 8397.65071 | VR | 23.08 | 0.05 | 600 | -0.07 |
| DCT | 8397.65784 | VR | 23.17 | 0.06 | 600 | 0.01 |
| DCT | 8397.66497 | VR | 23.16 | 0.06 | 600 | -0.01 |
| DCT | 8397.67207 | VR | 23.25 | 0.06 | 600 | 0.05 |
| DCT | 8397.67917 | VR | 23.19 | 0.05 | 600 | -0.05 |
| DCT | 8397.68630 | VR | 23.27 | 0.06 | 600 | 0.00 |
| DCT | 8397.69340 | VR | 23.39 | 0.07 | 600 | 0.07 |
| DCT | 8397.70050 | VR | 23.47 | 0.07 | 600 | 0.09 |
| DCT | 8397.70762 | VR | 23.54 | 0.07 | 600 | 0.10 |
| DCT | 8397.71473 | VR | 23.58 | 0.07 | 600 | 0.06 |
| DCT | 8397.72183 | VR | 23.76 | 0.08 | 600 | 0.16 |
| DCT | 8397.72895 | VR | 23.63 | 0.08 | 600 | -0.05 |
| DCT | 8397.73605 | VR | 23.80 | 0.09 | 600 | 0.05 |
| DCT | 8397.74315 | VR | 23.73 | 0.09 | 600 | -0.04 |
| DCT | 8397.75028 | VR | 23.90 | 0.10 | 600 | 0.15 |
| DCT | 8397.75738 | VR | 23.66 | 0.08 | 600 | -0.03 |
| DCT | 8397.76448 | VR | 23.67 | 0.07 | 600 | 0.06 |
| DCT | 8397.77161 | VR | 23.49 | 0.07 | 600 | -0.04 |
| DCT | 8397.77871 | VR | 23.42 | 0.06 | 600 | -0.03 |
| DCT | 8397.78581 | VR | 23.37 | 0.06 | 600 | -0.02 |
| DCT | 8397.79293 | VR | 23.32 | 0.06 | 600 | -0.01 |
| DCT | 8397.80004 | VR | 23.18 | 0.05 | 600 | -0.10 |
| DCT | 8397.80714 | VR | 23.20 | 0.05 | 600 | -0.04 |
| DCT | 8397.81426 | VR | 23.14 | 0.05 | 600 | -0.07 |
| DCT | 8397.82136 | VR | 23.09 | 0.05 | 600 | -0.09 |
| DCT | 8397.82846 | VR | 23.10 | 0.05 | 600 | -0.06 |
| DCT | 8397.83559 | VR | 23.13 | 0.05 | 600 | -0.02 |
| DCT | 8397.84269 | VR | 23.18 | 0.05 | 600 | 0.04 |
| DCT | 8397.84979 | VR | 23.08 | 0.05 | 600 | -0.06 |
| DCT | 8397.85692 | VR | 23.11 | 0.05 | 600 | -0.03 |
| DCT | 8397.87112 | VR | 23.13 | 0.06 | 600 | -0.01 |
| DCT | 8397.87824 | VR | 23.13 | 0.05 | 600 | 0.00 |
| DCT | 8397.88534 | VR | 23.06 | 0.05 | 600 | -0.07 |

## 3.2 Earlier HST Photometric Measurements

Table 4 lists resolved photometry of Manwë and Thorondor (magnitudes $M_M$ and $M_T$, respectively) previously obtained with the HST Wide-Field Planetary Camera2 (WFPC2) in the F606W band at seven different epochs from 2007 to 2013 (earlier observations on 2006 Jul 25 are excluded because no F606W-band data were obtained). Given the allocated orbits, only a few observations were obtained at each epoch, with each epoch spanning < 0.1 day. Grundy et al (2014) report magnitudes for Manwë and Thorondor derived from these data after transforming F606W to V band and averaging the results from each epoch. Here we present the measurements without transformation or averaging. For each epoch, we also list θ and the apparent separation, Δ, between Manwë and Thorondor. The listed magnitude errors are the variance with respect to the epoch mean. For the 2008 Sep 7 epoch, there is only a single measurement. We estimate an error of 0.1 mag consistent with the error for other epochs. An asterisk (*) indicates rejected measurements in the foregoing analysis (outliers, measurements with unusually large variance, or measurements made when Manwë and Thorondor were too close together to deconvolve reliably).

For each HST observation, we also list a pseudo magnitude for Manwë+Thorondor, M*, which is the F606W magnitude that would be observed for Manwë+Thorondor (unresolved) if the magnitude of Thorondor were fixed at its weighted mean value, $\langle M_T \rangle$ = 24.4 mag, determined from all the unrejected HST observations. The reason for constructing this derived signal will be discussed further below. Here we note that if Manwë's light curve is regular and periodic on a time scale of ~1 day, while Thorondor's brightness varies on a time scale comparable to the time between observing epochs (months to years), then a single rotational model will fit both the long-term variability of M* observed with HST in the years 2007 to 2013 and the more recently observed short-term variability we observe for the unresolved system with the SOAR, Gemini, and DCT telescopes.

**Table 4.** F606W-band Photometry of Manwë-Thorondor with HST

| UT Date | JD-2450000 (d) | $M_M$ (mag) | $M_M$ err (mag) | $M_T$ (mag) | $M_T$ err (mag) | M* (mag) | θ (deg) | Δ (arcsec) |
|---|---|---|---|---|---|---|---|---|
| 2007 Jul 25 | 4306.63168 | 23.71 | 0.05 | 24.31 | 0.09 | 23.25 | 0.85 | 0.184 |
| 2007 Jul 25 | 4306.64001 | 23.65 | 0.05 | 24.32 | 0.09 | 23.21 | 0.85 | 0.184 |
| 2007 Jul 25 | 4306.64765 | 23.75 | 0.05 | 24.34 | 0.09 | 23.28 | 0.85 | 0.184 |
| 2007 Jul 25 | 4306.68723 | 24.08* | 0.05 | 24.50 | 0.09 | 23.48* | 0.85 | 0.184 |
| 2007 Aug 26 | 4339.05946 | 23.64 | 0.10 | 25.01 | 0.69 | 23.21 | 0.21 | 0.036 |
| 2007 Aug 26 | 4339.06709 | 23.48 | 0.10 | 25.98 | 0.69 | 23.10 | 0.21 | 0.036 |
| 2007 Aug 26 | 4339.07473 | 23.42* | 0.10 | - | - | 23.05 | 0.21 | 0.036 |
| 2007 Aug 26 | 4339.08237 | 23.45* | 0.10 | - | - | 23.07 | 0.21 | 0.036 |
| 2008 Aug 4 | 4683.25518 | 23.37 | 0.05 | 24.58 | 0.07 | 23.01 | 0.67 | 0.180 |
| 2008 Aug 4 | 4683.26143 | 23.38 | 0.05 | 24.55 | 0.07 | 23.02 | 0.67 | 0.180 |
| 2008 Aug 4 | 4683.33990 | 23.27 | 0.05 | 26.06* | 0.07 | 22.94 | 0.67 | 0.180 |
| 2008 Aug 4 | 4683.34615 | 23.34 | 0.05 | 24.68 | 0.07 | 22.99 | 0.67 | 0.180 |
| 2008 Aug 20 | 4699.10379 | 23.38 | 0.11 | 24.62 | 0.25 | 23.02 | 0.35 | 0.283 |
| 2008 Aug 20 | 4699.11004 | 23.28 | 0.11 | 24.80 | 0.25 | 22.95 | 0.35 | 0.283 |
| 2008 Aug 20 | 4699.18851 | 23.26 | 0.11 | 24.68 | 0.25 | 22.93 | 0.35 | 0.283 |
| 2008 Aug 20 | 4699.19476 | 23.49 | 0.11 | 25.18 | 0.25 | 23.10 | 0.35 | 0.283 |
| 2008 Sep 7 | 4717.08724 | 23.90 | 0.10 | 24.57 | 0.10 | 23.37 | 0.08 | 0.329 |
| 2008 Oct 26 | 4766.30316 | 23.65 | 0.08 | 24.80 | 0.25 | 23.21 | 1.00 | 0.078 |
| 2008 Oct 26 | 4766.31150 | 23.68 | 0.08 | 24.86 | 0.25 | 23.23 | 1.00 | 0.078 |
| 2008 Oct 26 | 4766.31914 | 23.56 | 0.08 | 24.44 | 0.25 | 23.15 | 1.00 | 0.078 |
| 2008 Oct 26 | 4766.32747 | 23.76 | 0.08 | 25.04 | 0.25 | 23.28 | 1.00 | 0.078 |
| 2013 Nov 20 | 6617.16260 | 23.53 | 0.08 | 24.39 | 0.05 | 23.13 | 1.21 | 0.223 |
| 2013 Nov 20 | 6617.16588 | 23.70 | 0.08 | 24.29 | 0.05 | 23.24 | 1.21 | 0.223 |
| 2013 Nov 20 | 6617.23032 | 23.70 | 0.08 | 24.40 | 0.05 | 23.24 | 1.21 | 0.223 |
| 2013 Nov 20 | 6617.23360 | 23.62 | 0.08 | 24.38 | 0.05 | 23.19 | 1.21 | 0.223 |

### 3.3 Period Determination and Sinusoidal Light-curve Fitting

Here we determine the rotation period P by simultaneously fitting a periodic function, M(t), to all the photometry listed in Table 3 with the exception of the first DCT epoch (2017 Nov 8) which were collected in bad observing conditions, as noted above. We also fit the un-rejected combined HST photometry of Manwë+Thorondor listed in Table 4, M*. As discussed above, we construct M* assuming a brightness for Thorondor fixed at its mean HST brightness. Our assumption, here, is that the light curve of Manwë is regular and periodic over the entire time period spanned by the observations, while the variability of Thorondor is on a time scale much longer than any given observation epoch. The constructed HST photometry would therefore have the same periodicity and shape as the average light curve observed over a single epoch from the ground.

Following Rabinowitz et al (2014), we express M(t) as

$$M(t) = H_j + \beta\theta(t) + F(t) \quad (1)$$

where $H_j$ is the mean reduced magnitude in band j at phase angle $\theta = 0$ (i.e. the absolute magnitude in each band), t is the light-travel corrected observation time, $\beta$ parametrizes a linear dependence on $\theta$. Function F(t) is a single harmonic sinusoid with harmonic amplitude A and phase $\phi$,

$$F(t) = (A/2)\sin(\pi\,[2\phi + 4\omega(P,t)]) \quad (2)$$

where $\omega(P,t)$ is the rotational phase, $\omega(P,t) = (t-t_o)/P$, referred to an arbitrary epoch, $t_o$. There are two peaks per rotation, corresponding to the light curve of a rotating ellipsoid.

The first row of Table 5 lists the minimum $\chi^2$, the number of degrees of freedom, N (239 observations less 7 fitted parameters), and the best-fit values and uncertainties for P, $\beta$, A, $\phi$, the absolute magnitudes $H_g$, $H_r$, and $H_{F606W}$, and g-r. Here, we have assumed $t_o$= JD 2544300.0. Here we use the $\chi^2$ minimization procedure described in detail by Rabinowitz et al. 2014. We also used Monte Carlo simulations to determine the 68% confidence intervals of the associated parameters. These simulations preserve the observation times and measurement errors, but use the best-fit light curve to model the magnitude values. These confidence intervals are valid if the best-fit light curve is a good representation of the observed signal (i.e. $\chi^2 \sim 1$). The second row of Table 5 refers to an improved fit after making additional adjustments to the data. We discuss the differences in the resulting parameters and the significance of the estimated errors further in Sec. 3.4.

Note that the resulting confidence intervals were determined after adding a systematic error of 0.025 mag in quadrature to the measurement errors listed in Tables 3 and 4. This accounts for measurement errors associated with the residual signal left by incompletely subtracted stars and galaxies along the target trajectory. Adding this offset affects only the confidence limits we find for the fitted parameters, not their best-fit values. The listed errors for $H_g$, $H_r$, and $H_{F606W}$ do not include the calibration zero-point uncertainties (~ 0.02 mag).

**Table 5. Best-Fit Light Curve Parameters for Manwë and Thorondor Combined**

| $\chi^2$ | N | P (h) | β (mag/deg) | A (mag) | ϕ | $H_g$ (mag) | $H_r$ (mag) | $H_{F606W}$ (mag) | g-r (mag) |
|---|---|---|---|---|---|---|---|---|---|
| 3.02 | 232 | 11.88190 | 0.000 | 0.480 | 0.310 | 7.66 | 6.85 | 6.62 | 0.81 |
| | | ±0.00005 | ±0.01 | ±0.02 | ±0.02 | ±0.02 | ±0.01 | ±0.02 | ±0.02 |
| 2.37 | 232 | 11.88185 | 0.050 | 0.480 | 0.250 | 7.62 | 6.79 | 6.61 | 0.83 |
| | | ±0.00005 | ±0.01 | ±0.02 | ±0.02 | ±0.02 | ±0.01 | ±0.03 | ±0.02 |

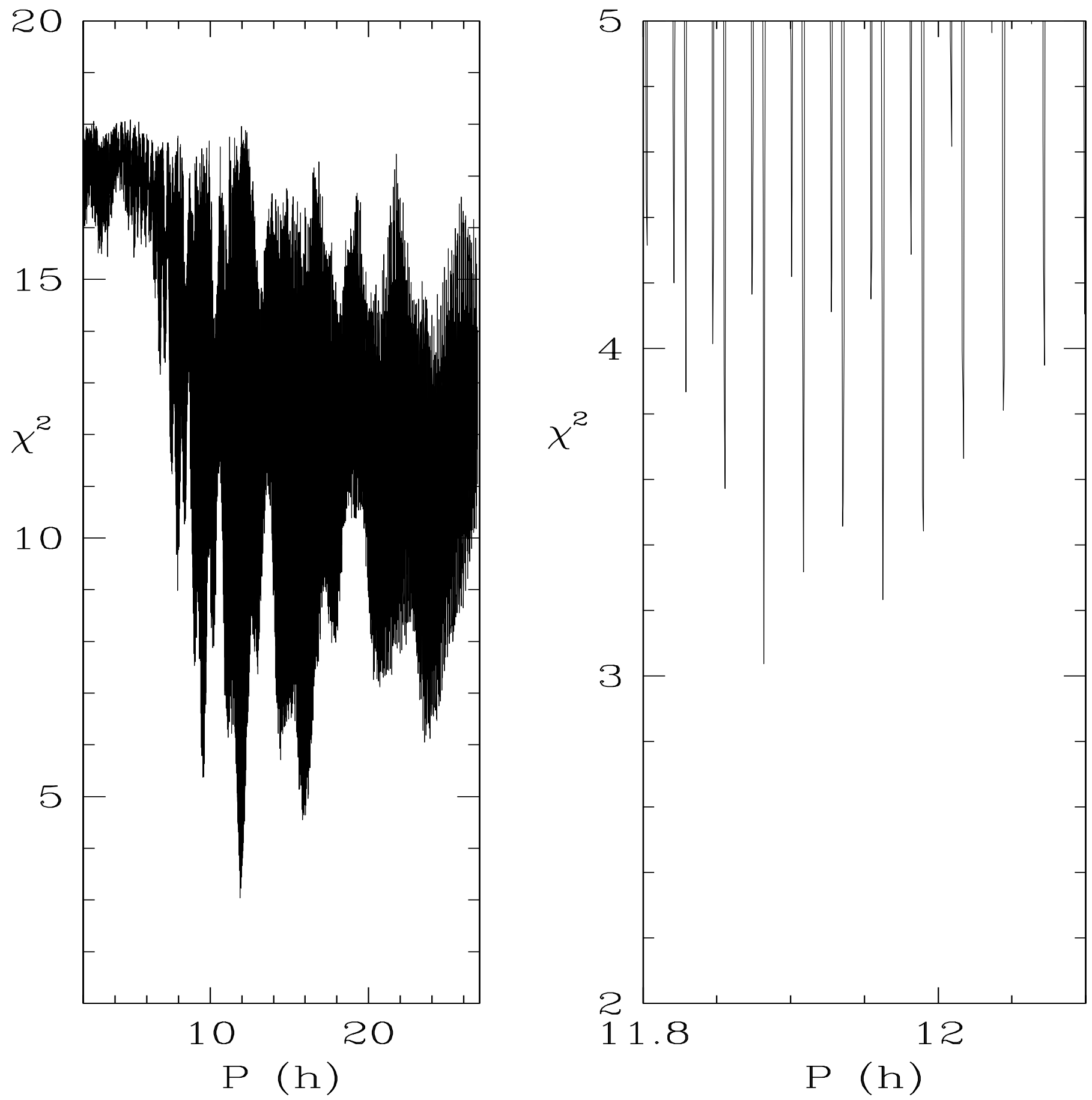

**Figure 1**. Reduced $\chi^2$ vs period P for the best-fit double-peaked light curve, simultaneously fit to the SOAR, Gemini, and DCT observations of Manwë-Thorondor and to earlier HST observations of Manwë after adjustment to include a contribution from Thorondor (see text for details) . Left side shows full range of reduced $\chi^2$ over full range of the period search. Right side shows values near minimum reduced $\chi^2$. See Table 5 for best-fit parameters.

Figure 1 shows the minimum reduced $\chi^2$ we find versus P, with the best fit occurring at P = 11.88190 h with reduced $\chi^2$ = 3.02 (239 observations and 7 fitted parameters). Note that there are many other $\chi^2$ minima near the best fit (see right side of Fig. 1). These are aliases with separation ΔP ~ 0.027 h, equivalent to one extra period in the long interval (438 days) between the two data sets most critical for fixing P (Gemini in 2017 Jul and DCT in 2018 Oct). The most significant secondary minimum occurs at P = 11. 96254 h with reduce $\chi^2$ = 3.232. From Monte Carlo simulations, we find that these aliases yield a lower $\chi^2$ than the true period less than 1% of the time. Hence, we reject this secondary minimum at the 99% confidence level.

3.4 Best-fit Sinusoid Compared to Observations

Figure 2 shows our best-fit sinusoidal light curve (solid line) as a function of rotational phase superimposed on the reduced observations. Here the observations are separated vertically by the date of observation, starting with the constructed HST values (M*, as described above), followed by the SOAR, Gemini, and DCT nights. See the figure legend for symbol explanations. Note that the bottom light curve is a second representation of the 2nd DCT epoch (2018 Oct 6) showing the assumed effect of increasing the brightness contribution from Thorondor (see discussion below). Also note that g band data (red squares) are shifted relative to r-band data (black squares) by subtracting the fitted value for g-r listed in Table 5. All error bars include an added systematic error of 0.025 mag to account for residual, un-subtracted background flux along the target path in the images. The dashed lines show the nominally expected diminution from predicted mutual events at the relevant epochs, high-lighted in green where the diminution exceeds 10% (discussed further below).

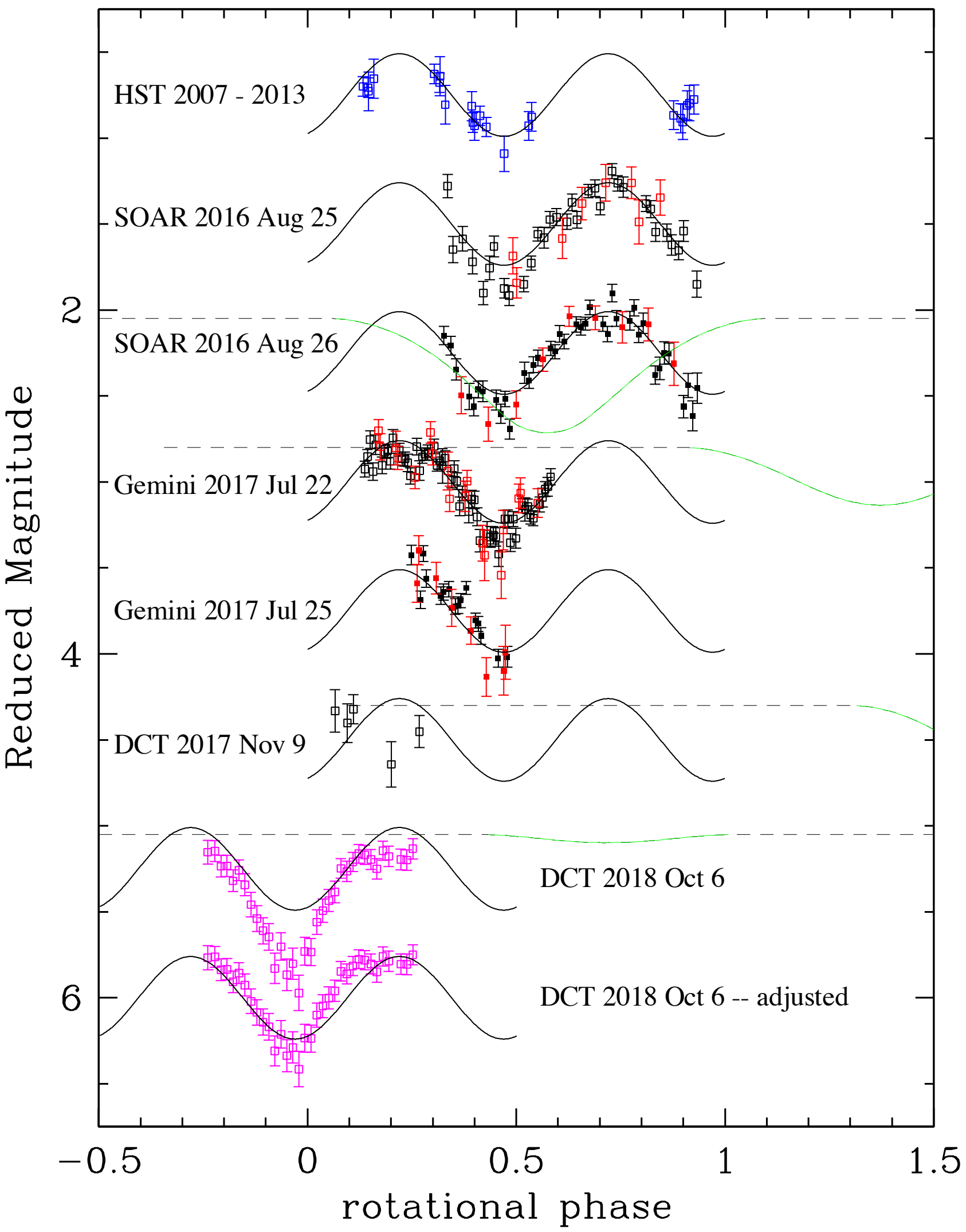

**Figure 2.** Reduced magnitude versus rotational phase for Manwë-Thorondor as measured (from top to bottom) with HST in the F606W band (assembled from six different epochs 2007 Jul 25 to 2013 Nov 20 -- see M* in Table 4 and text for details), SOAR in the g and r bands on 2016 Aug 25 and again on Aug 26, Gemini South in the g and r bands on 2017 Jul 22 and again on 2017 Jul 25, DCT in r band on 2017 Nov 9, DCT in the VR band on 2008 Oct 6 before alteration, and the same DCT VR-band data after alteration (see text for details). Blue, black, red, and magenta squares represent the F606W, r, g, and VR bands, respectively. All observations have been adjusted to remove their fitted dependence on solar phase angle. The solid black line shows the best fit double-humped sinusoid (see Table 5 for model parameters). Each light curve has been shifted arbitrarily in the vertical direction for clarity, with the g band shifted relative to r by subtracting the fitted value for g-r. The error bars show the statistical measurement uncertainty plus a systematic error of 0.025 mag added in quadrature. The dashed lines show the nominally predicted diminution from Thorondor eclipsing and occulting Manwë (see Table 2 and Grundy et al. 2014) where the dependence on rotational phase is computed assuming the period and phase parameters listed in Table 5. Green color indicates the phase range where the diminution exceeds 10%. Note that rotational modulation is not modelled in these event predictions, and the uncertainty in the timing of these mutual events is equivalent to a full rotational phase. The duration and depth of the events are also uncertain by a factor ~2 given the uncertainty in the diameters of the two bodies.

It is clear by inspection of Fig. 2 that our best-fit, single-harmonic sinusoid accounts for most of the signal variation we see from night to night (~0.5 mag peak to peak) with the exception of the second-epoch of DCT observations. We discuss the DCT observations further, below. Excluding these DCT data, the most significance deviations from the fit occur on the first Gemini night (2017 Jul 22) over the phase range $\omega = 0.1$ to 0.3. There the sinusoidal fit comes to gradual maximum whereas the observed light curve flattens and then dips by ~0.2 mag (near $\omega = 0.25$) relative to the fit. These deviations occur in both the g and r band, and are not an artifact due to un-subtracted background sources in the field along the target trajectory. The second peak (near $\omega = 0.7$, covered only with the SOAR observations) does not appear to flatten significantly relative to the sinusoid. However, these data have larger scatter than the Gemini observations. There could be some flattening in the second peak that is unresolved.

With respect to the 2nd-epoch DCT data, the best-fit sinusoid phases well with the observations, but the DCT data have a larger peak-to-peak variation (~0.7 mag, compared to 0.5 mag for the model) and a fainter average signal (by 0.15 mag). The DCT light curve also flattens significantly from $\omega = 0.07$ to 0.25 relative to the sinusoid. This mirrors a similar flattening in the Gemini light curve that occurs over the same $\omega$ range. On the assumption that the disparities in the DCT observations are the result of a diminished flux from Thorondor relative to earlier epochs (as might be expected from the ~0.7 mag variability observed by HST), we computed the hypothetical light curve we would have measured if the flux from Thorondor were increased to its nominal value.

The resulting adjusted light curve appears at the bottom of Fig. 2. Here we assume that the actual r-band magnitude of Thorondor at the time of the DCT observations was 25.4 mag (one magnitude fainter than average F606W magnitude observed by HST). We brighten Thorondor by one magnitude to compute the adjusted light curve. It is apparent that the adjusted light curve is now in significantly better agreement with the best-fit sinusoid. The average brightness and the peak-to-peak variation of the two curves are nearly matched. Refitting the observations with this adjustment improves the reduced $\chi^2$ from 3.02 to 2.39. Significant differences remain, however. Between $\omega = 0.0$ and 0.2, the adjusted light curve rises more steeply (2.9 mag/period) than the sinusoid (2.0 mag/period). This phase range was not recorded at earlier epochs, so we cannot verify if this inconsistency was present at earlier epochs.

The second row of Table 5 lists the new values for the light-curve parameters that result from fitting the adjusted observations. The changes are insignificant for P, A, $H_{F606W}$, and g-r. Given the improved $\chi^2$, this validates our uncertainty estimates for these parameters. There is a small, but significant change in β (from 0.00 to 0.05 mag/deg), which is coupled to small decreases in $H_g$ and $H_r$ (by 0.04 mag). These differences are a likely indication of our fitting uncertainty, which are larger than we have calculated from Monte-Carlo estimates.

From the above comparisons we can conclude that the combined light curve for Manwë and Thorondor is periodic, with the peaks and valleys consistently occurring at the same rotational phase from the earliest HST observations (2007 Jul) to the latest ground-based observations (2018 Oct). For most epochs the light curve is sinusoidal to 1st order. The significant differences seen in the latest DCT data could be explained by Thorondor's long-term variability. In section 3.6 we present a multi-component model to better model the over-all shape of the observed light curves.

### 3.5 Possible Mutual Event Detections and Non-detections

Below we examine the light curves measured at each observed epoch to determine if a mutual event occurred. As will become evident, it is likely we detected none of the predicted events, or that our detections were at most marginal. The reasons could be that the time span of our individual light curves did not sufficiently cover the uncertainty window of the event times. More interestingly, it could be that the nominal event predictions, while based on the most likely assumptions, were unrealistic because of unusual shapes or sizes for Manwë and Thorondor.

*Epoch 1 (SOAR data)*. As shown by the dashed lines in Fig. 2, we would have detected a significant diminution in brightness of Manwë-Thorondor on the 2nd SOAR night if the maximum diminution (0.66 mag), the event duration (~12 hour or 1 full rotational phase), and the event mid time were as predicted. Even if the event mid time were early by 8 hours or late by 10 hours, the expected diminution would exceed 0.1 mag at the beginning or end of the observational window, respectively. However, given the close similarity of our light curves from both SOAR nights, it is clear that such a significant diminution did not occur. Taking into account only the uncertainty in the timing of the event, our non-detection had a likelihood of ~40%.

*Epoch 2 (Gemini data)*. Based on the above outcome, we planned our Gemini observations to detect a subsequent inferior event on the assumption that the actual mid-time would be outside the ±1 sigma uncertainty window. Unfortunately, bad weather prevented us from scheduling our second night optimally. We therefore have no coverage in epoch 2 to verify a late event time. However, the first Gemini night would have overlapped the nominally predicted event (with the diminution exceeding 0.1 mag) if the event predictions were early by at least 10 hours.

Comparing the phased light curves from our two Gemini nights to each other and to the earlier SOAR nights, it again appears that no significant diminution occurred. As discussed in the previous section, there are some significant deviations on the first Gemini night relative to the best-fit sinusoid. Some of these deviations (e.g. the dip at $\omega$ =

0.25, or else the deeper than expected minimum near $\omega = 0.45$) might result from a mutual event if both maximum diminution and duration were smaller than expected. We address these deviations again after presenting an improved rotational model for the system in Sec 3.6. The main conclusion is that the mutual event did not occur earlier than expected if the depth and duration were as nominally predicted. This leaves open the possibility the event was late.

*Epoch 3 (DCT 2017 Nov 9).* The DCT observations scheduled 2017 Nov 9 would have detected a nominally predicted inferior event if it were early by 20 hours (2 sigma). Given the poor weather conditions and sparse observations, no conclusions can be drawn. The observations are marginally consistent with the best-fit sinusoid at the 1-sigma level.

*Epoch 4 (DCT 2018 Oct 6).* The DCT observations scheduled on 2018 Oct 6 would have covered a nominally predicted inferior event if it occurred earlier than predicted by 5 to 14 h. However, the expected depth for this event was small (0.05 mags). As discussed above, the light curve we observed on this night deviates significantly from the best-fit sinusoid, with deeper minimum and a lower average brightness. If the event depth were much larger than predicted, some of these variances might be the outcome of the mutual event. To be consistent with the negative detections at earlier epochs discussed above, this would imply that the model used to predict the mutual events is grossly in error, with the events predicted at earlier epochs not occurring at all, or occurring with much lower diminution than expected. We return to this discussion later, after presenting an improved rotational model.

### 3.6 A Multi-component Model

To better fit both the short- and long-term variations in the combined light curve of Manwë and Thorondor, we construct a model to predict the variability for a multi-component system. Here we assume Manwë is a contact binary, rotating with period, $P_M$, and composed of two equal-sized lobes with matching, uniform albedos. This system is orbited by Thorondor, an independent body rotating with period $P_T$. The model determines the relative positions of the Sun, Earth, and all other bodies in the system and computes the total brightness variations caused by the mutual occultations and eclipses. In this paper, we assume a Lambertian scattering law for all surfaces.

As a starting point, we assume Manwë is a barbell-shaped contact binary composed of two equal-sized, elongated lobes connected by a narrow bridge (see Fig. 3). As shown by Gnat and Sari (2010) and again by Descamps (2015), this is the shape expected for a fluid body with normalized rotational velocity, $W = [3\pi/\rho G]^{1/2}/P \sim 0.30$, where $\rho$ is the density and G is the gravitational constant. As W approaches the critical value 0.2815, the narrow bridge connecting the two lobes vanishes and the hydrostatic solution becomes a pair of disconnected tidally-locked bodies. Descamps uses shape models with W ranging from 0.30 to 0.32 to fit the light curves of the suspected contact-binary (139775) 2001 QG298 (Sheppard & Jewitt 2004), and the main-belt contact-binaries (624) Hektor and (216) Kleopatra. Based on this earlier work, we choose the shape model corresponding to W = 0.31. With the rotation period we measure for Manwë (P = 11.88 h), this requires $\rho$ =

0.80 g/cm$^3$. This density value is near the nominal value measured for KBOs of comparable size (Stansberry et al. 2008), near the value (0.75 g/cm$^3$) estimated by Grundy et al. (2014) for the combined system bulk density, and within the range of values estimated for other likely contact binaries (Thirouin & Sheppard 2017, 2018). Hence, this shape model is a good starting point for modeling Manwë's light curve.

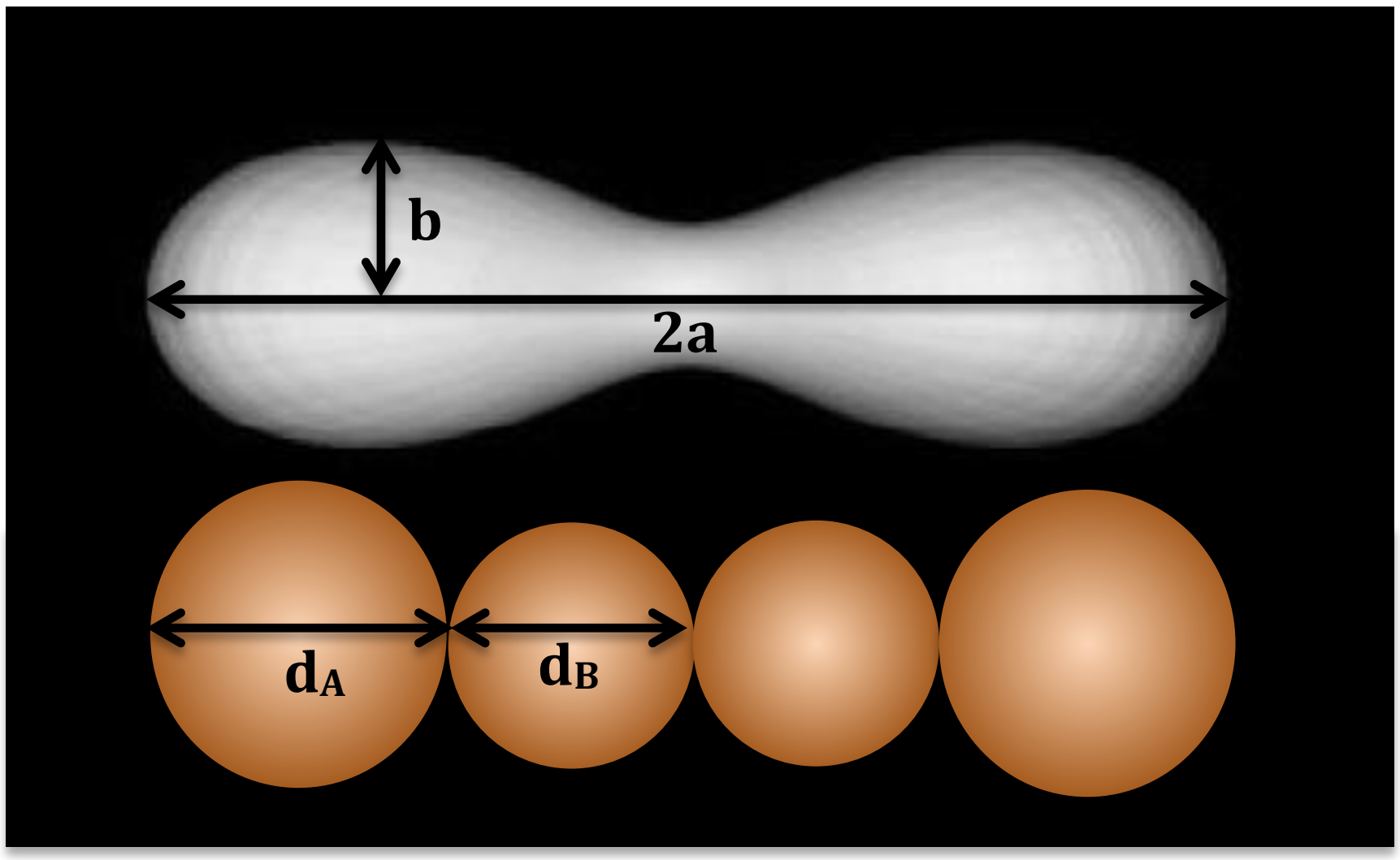

Figure 3. Shape models for Manwë used for the multi-component light curve calculations presented in this paper. The barbell model (grey-shaded figure, with long axis, 2a, and maximum perpendicular radius, b) is the shape calculated by Descamps (2015) for a strengthless body with normalized rotational velocity [3π/ρG]1/2/P = 0.31. To simplify our light curve calculations, we represent each lobe of this figure by a pair of contacting spheres (orange shading) with respective diameters dA = 2b and dB = a – 2b.

For simplicity and to speed our computations, we approximate each lobe of the barbell shape calculated by Descamps by a pair of contacting spheres, A and B, with diameters $d_A$ and $d_B$, aligned with the long axis of the barbell. To set their relative scale, we refer to the shape parameters a and b (the half the length of the barbell and its maximum cross-sectional radius, respectively) given by Table 1 of Descamps (2015). Scaled to the radius of a sphere with the same volume as the barbell, a = 2.284 and b = 0.694. We set $d_A$ to a value slightly less than the maximum cross-sectional diameter (1.8 x b) and $d_B$ to span the gap between sphere A and the barbell's midpoint (a – 1.8 x b) . This yields $d_B/d_A$ = 0.830. At maximum brightness viewed perpendicular to its spin axis, Manwë then has a surface area equivalent to a single sphere with diameter $d_M = d_A[2(1+(d_B/d_A)^2)]^{1/2}$ = 1.838$d_A$ and a volume equivalent to a sphere of diameter $d_{MVOL} = d_A[2(1+(d_B/d_A)^3)]^{1/3}$ = 1.465$d_A$ = 0.797$d_M$. With this shape, we are able to produce a light curve differing by less than 10% from the one generated by Descamps (2015) for the case W = 0.32, assuming the same scattering law.

To model Thorondor's contribution to the long-term variations of the light curve, we assume a spherical body with diameter, $d_T$, and a brightness that oscillates sinusoidally with period $P_T$ (single-peaked) and amplitude $A_T$ mag. This variation could be due to albedo variations across the surface (e.g. a dark surface feature rotating into and out of view). We further assume Thorondor's brightest visible hemisphere has an average albedo matching Manwë's albedo. This would be consistent with the matching colors observed for Manwë and Thorondor (Grundy et al. 2014) and for most KBO binaries (Benecchi et al. 2009). We recognize that Thorondor's shape is likely elongated and the dominant contribution to Thorondor's brightness variations. This is an important consideration because in mutual events with Manwë, Thorondor's shape is a significant factor affecting the short-term time dependence of the light curve. We therefore do not expect our simplified multi-component model to accurately predict light curves during these particular events. At all other times, however, the shape we assume for Thorondor is less important. We assume a spherical shape only as a convenience to calculate Thorondor's peak brightness relative to Manwë. We revisit this assumption later. As long as the resulting variability is slow, periodic, and approximately sinusoidal, our model predictions will be valid. To avoid the complication of modeling the mutual events, we assume that no significant mutual event occurred during the observations, consistent with the conclusion of the previous section.

Despite our simplifying assumptions, the list of input parameters we need to completely specify our multi-component model are numerous (see Table 6) : diameters $d_T$ and $d_M$, the heliocentric orbit of the tertiary system, the orbital elements for Thorondor's motion about Manwë's center of mass, Manwë's rotation period, $P_M$, and the orientation of Manwë's equator (ascending node, $\Omega$, and inclination, i, referenced to the J2000 equatorial frame) and parameters $P_T$ and $A_T$ characterizing Thorondor's rotational light curve. We must also specify a reference epoch ($\varepsilon_T$) for Thorondor's rotational phase (chosen to be the date of the most recent minimum) and the rotational angle ($\mu$) of Manwë's long axis relative to the ascending node of Manwë's equator at a given epoch.

Fortunately, most of these parameters are known or well constrained. We refer to the JPL Horizons website[2] for the heliocentric orbital elements. We refer to Grundy et al. (2014) for Thorondor's orbit about Manwë, but we adjust the orbital period within its range of error (adding 0.020 d) so that no mutual events occur during any of the epochs covered by the observations. We fix $P_M$ at the value we initially determine by fitting a sinusoid to the observed light curves (see Table 5). We set $\varepsilon_T$ = JD 2458398.0, coinciding with the second DCT epoch, when we expect Thorondor to be near minimum brightness. We can also set $d_T$ and $d_M$ by adopting the nominal respective values, 92 and 160 km, estimated by Grundy et al (2014). However, we scale these values up slightly by a factor 1.16 so that $d_M$ = 188 km and $d_T$ = 108 km. We then have an average density for the system, $\rho = m_{sys}/[(\pi/6)*([0.839d_M]^3 + d_T^3)] = 0.80$ g/cm$^3$, matching the density required by our shape model.

There are nine free parameters that remain to be determined by fitting our multi-component model to the observations: the rotational orientation of Manwë ($\Omega$, i, and $\mu$), the light-curve parameters for Thorondor ($P_T$, $A_T$), the absolute magnitudes of the combined system in the observed pass bands ($H_g$, $H_r$) and the separate absolute magnitudes for Manwë and Thorondor in the F606W band pass ($H_M$ and $H_T$). We reference each absolute magnitude to the fixed epoch, $t_1$ = JD 2457625.0, chosen to be a time when Manwë and Thorondor are both near the peak of their separate light curves. We note here that $\Omega$ and i are coupled. All pairs of values that yield the same angle, $\alpha_M$, between Manwë's rotational pole and the Manwë-Earth direction (or anti-direction) at a given epoch will yield the same predicted magnitudes. Because the Manwë-Earth direction changes only by ~10 deg during the total time spanned by the observations (including HST), $\alpha_M$ is nearly constant. The geometry provides little leverage for decoupling $\Omega$ and i.

[2] https://ssd.jpl.nasa.gov/?horizons

To obtain the fit, we use the same $\chi^2$ minimization procedure described above (see Sec. 3.3). Instead of using a sinusoid to described the model light curve, F(t), in Eq. (1), we substitute the light curve computed using our multi-component model. As before, we simultaneously fit all the observations listed in Tables 3 and 4, but we now separate the HST photometry for Manwë and Thorondor into two independent data sets. For each value of F(t), we therefore calculate offsets $\delta F_M(t)$ and $\delta F_T(t)$, which we add to F(t) to yield the predicted magnitudes for Manwë and Thorondor, respectively. We also normalize F(t) differently than before, defining $F(t_1) = 0$ mag at epoch, $t_1$. In our earlier sinusoidal fit, we normalize the model light curve so that F(t) = 0 mag at its average value. This leads to a relative offset for the computed values of $H_g$, $H_r$, and $H_{F606W}$. Also, because of this normalization, it is only the ratio of model parameters $d_T$ and $d_M$ that affects the $\chi^2$ values of our fits. If our shape model is correct, however, then we know the density of the system. The sizes are then constrained by the known total mass.

Table 6 lists the resulting best-fit values for the free parameters and for the fixed parameters discussed above. The reduced $\chi^2$ of the fit is 2.0 for 255 observations with 9 free parameters. The residuals to the fit are listed in the last column of Table 3. This is a significant reduction compared to our sinusoidal fit ($\chi^2 = 3.02$). Figure 4 shows the best-fit model light curve compared to the observations. Except for the top curve, these are the same observations we present in Fig. 2, again plotted versus rotational phase, $\omega(P,t) = (t - t_0)/P_M$, with the value for $t_0$ listed in Table 6. The top curve shows the residuals to the fit of the magnitudes for Manwë and Thorondor separately measured with HST. Figure 5 shows the residuals for all the measured magnitudes plotted versus $\theta$ (averaged over bins of width $\Delta\theta = 0.01$ deg).

**Table 6. Multi-component Model Parameters**

| | | |
|---|---|---|
| $d_M$ | 188 km | diameter of sphere with equivalent area as Manwë |
| $d_{MVOL}$ | 150 km | diameter of sphere with equivalent volume as Manwë |
| $d_T$ | 108 km | diameter of Thorondor |
| $m_M$ | $1.41 \times 10^{18}$ kg | mass of Manwë |
| $m_T/m_M$ | 0.38 | upper limit to Thorondor/Manwë mass ratio[a] |
| $p_M$ | 0.06 | albedo of Manwë |
| $p_T$ | 0.09 | albedo of Thorondor |
| $\beta$ | 0.0 mag/deg | linear dependence on solar phase angle |
| $P_M$ | 11.88190 h | rotation period of Manwë |
| $t_0$ | 2457625.8 | reference Julian date for absolute magnitudes |
| $\rho$ | 0.80 g/m$^3$ | Average density of the Manwë/Thorondor system |
| $\varepsilon_T$ | 2458398.0 | Julian date of minimum brightness for Thorondor |
| $\Omega$ | 192.0 deg | ascending node of Manwë's equator[b] |
| i | 27.0 deg | inclination of Manwë's equator[b] |
| $\mu$ | 7.0 deg | rotation angle of Manwë's long-axis[c] |
| $\langle\alpha_M\rangle$ | 74.0 deg | average angle between Manwë's spin and Manwë-Earth direction |
| $P_T$ | 309.03 d | rotational period of Thorondor (single-peaked) |
| $A_T$ | 0.55 mag | light-curve amplitude of Thorondor (peak-to-peak) |
| $H_g$ | 7.48 mag | Absolute g-band magnitude of Manwë/Thorondor[d] |
| $H_r$ | 6.66 mag | Absolute r-band magnitude of Manwë/Thorondor[d] |
| $H_M$ | 6.94 mag | Absolute F606W-band magnitude of Thorondor[d] |
| $H_T$ | 7.73 mag | Absolute F606W-band magnitude of Thorondor[d] |

Table Notes: [a]assuming equal density, $\rho$, for Manwë and Thorondor; [b]J2000 equatorial coordinates; [c]relative to the ascending node of Manwë's equator at JD 2454300.0; [d]at epoch $t_0$;

Figure 4 shows that the new light-curve fit is in significantly better agreement with the observations than the sinusoidal fit shown by Fig. 2. The model now matches the maximum of the light curve between rotational phases 0.1 and 0.4 (Gemini 2017 Jul 22 and DCT 2018 Oct 6) while simultaneously fitting the V-shaped minima appearing in the Gemini and DCT light curves. Taking into account the long-term rotational variability of Thorondor, the model is now able to match the relatively deeper minimum and overall decrease in brightness for the combined system observed by DCT on 2018 Oct 6. There are some observations of Thorondor near rotational phase ~1.0 that have large scatter, and mostly fall well below the model prediction (magenta points in top curve). However, all these outliers correspond to the epochs when Manwë and Thorondor are separated by less than 0.08 asec (JD 2454399 and 2444766), making the deconvolution of the separate fluxes of Manwë and Thorondor unreliable. Because these observations have large errors, they have negligible influence on the model fit.

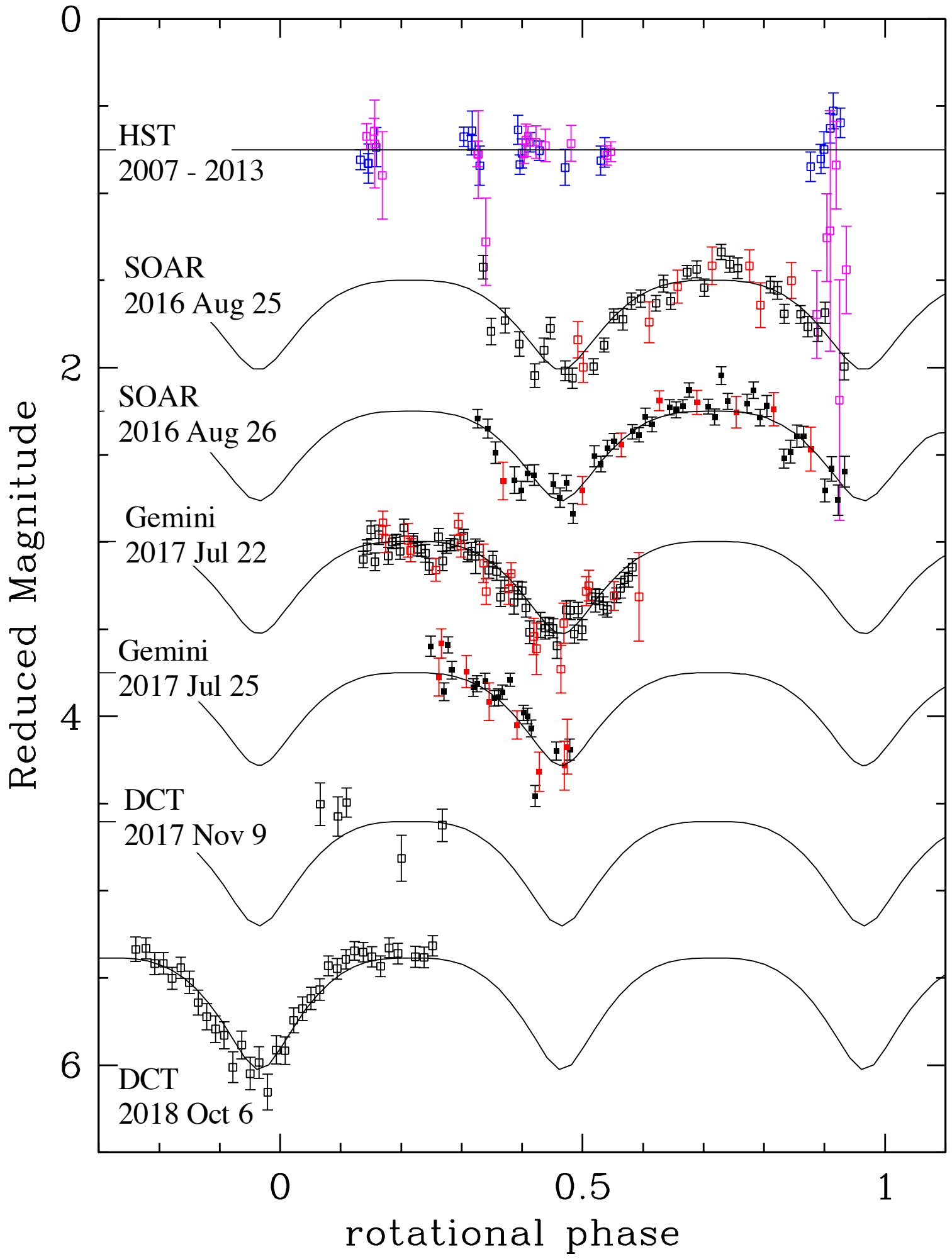

**Figure 4.** Reduced magnitude versus rotational phase for Manwë-Thorondor. These are the same data represented in Figure 2 (see legend), but now compared to a model light curve in which Manwë is a contact binary composed of two equal-sizes lobes, each constructed as a pair of spheres (see Fig 3), and Thorondor has a rotational light curve with a ~300-d period and 0.55-mag amplitude. The rotation period for Manwë matches the best fit value listed in Table 5. The orientation of the Manwë's spin axis and Thorondor's rotation period are optimized to best fit the observations (see Table 6). The assumed orbital period for Thorondor has been adjusted (within its measured uncertainty range) so that all the mutual events occur outside the time ranges of each night's observations. Manwë has a surface area equivalent to a sphere with diameter 186 km, and Thorondor has a diameter of 107 km. A Lambertian scattering model is assumed for all surfaces. Note that the top curve now shows the residuals to the model fit of the magnitudes for Manwë (blue) and Thorondor (magenta) separately measured with HST in the F606W band. A phased light curve is not shown for these data because the light curve amplitude and shape are variant over the long time span of the HST observations. Also note that the bottom light curve is a duplication of the un-adjusted, 2nd-epoch DCT data shown second to the bottom in Fig 2.

Note that we do not make an exhaustive search of parameter space to obtain this fit. We first sample the parameters on a coarse grid, and then search a finer grid only near the best best-fit value of the coarse search. The particular orientation we obtain for Manwë's equator (Ω=192, i=27 deg) corresponds to a pole orientation with $\alpha_M$ ranging from 69.0 to 78.9 deg over the time spanned by the observations in Tables 3 and 4. The pole's angle with respect Thorondor's orbital pole is 12.5 deg. However, as we discuss above there are other pole orientations that have consistent values for $\alpha_M$ and yield fits with only slightly higher $\chi^2$. For example, we obtain $\chi^2 = 2.25$ with $\Omega = 339$ deg and i = 30 deg. The corresponding pole has a very different angle with respect to Thorondor's orbit (55.5 deg) while the $\alpha_M$ range (77.6 to 78.9) is not significantly changed.

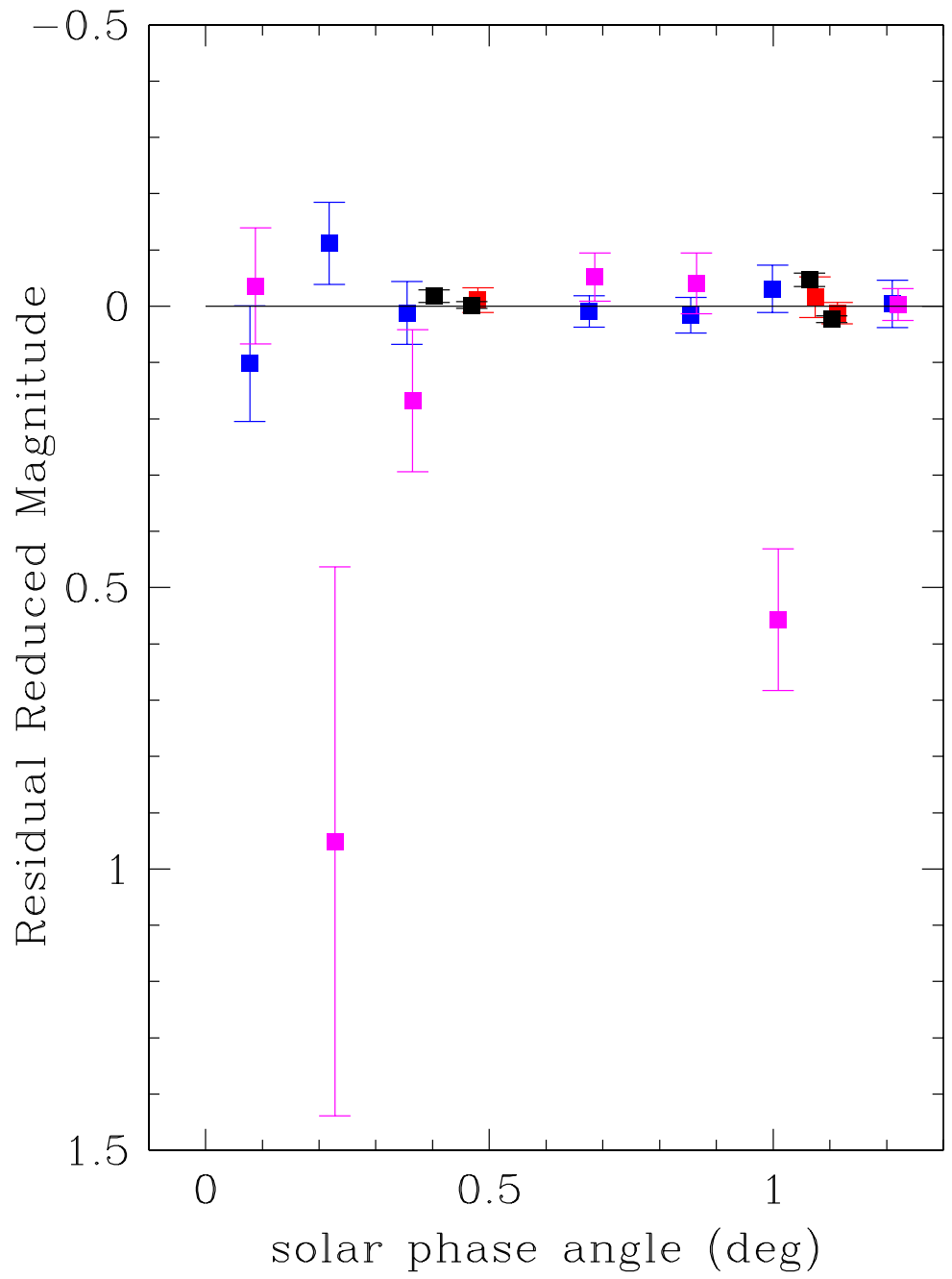

**Figure 5.** Averages of the residual reduced magnitude versus solar phase angle for Manwë-Thorondor after subtracting the best-fit multi-component model (see Fig. 4 and Table 6). See Fig. 4 for symbol explanations.

Regarding the best-fit value for $P_T$, we find alternate solutions with $P_T$ =3.2359 and 331.13 d that yield nearly the same $\chi^2$ as the best fit at $P_T$ = 309.03 days. The shorter period is very close to an integer ratio (13/2) with Manwë's rotation period (0.495079 d), while the longer period is nearly an integer multiple (3) of Thorondor's orbital period (110.176 d). Both of these alternative values are likely aliases of the best-fit period generated by sampling bias. When fitting simulated observations that have the same observation dates and error bars as the real data but with magnitudes generated using our multi-component model (assuming $P_T$ = 309.03 days), we find these same aliases. In 94/100 simulated trials, however, the true period yields the best fit. Assuming our model is a realistic description of the system, we have 94% confidence that our fit has identified the true period. If the actual light curve for Thorondor is not sinusoidal, other period solutions are possible.

## 4. DISCUSSION

### 4.1 Color Variation

Table 5 list the best fit color, g-r = $H_g - H_r$ = 0.81 ± 0.03 mag we obtain for Manwë-Thorondor assuming a sinusoidal fit. Our multi-components fit yields nearly the same value (0.82). This is consistent with the value, g-r = 0.85 ± 0.06 mag, reported by Sheppard (2012). From inspection of Figure 2 or 4, we can also conclude that there is no significant deviation between the g- and r-band light curves. At the level of our g-band measurement errors (5 to 10%), there is no dependence of color on rotational phase. Because Manwë's rotational variability is the dominant contribution to the light curve variability on short time scales, we can conclude that Manwë's hemispherically-averaged color is uniform with rotational longitude, although there might be a dependence on latitude. To the extent that color is a tracer of albedo (as it is on Pluto; Olkin et al 2017), this also indicates a uniform hemispherically-averaged albedo.

### 4.2 Solar Phase Angle Dependence

We derive an unusually flat slope to the solar phase curve for both Manwë and Thorondor from our sinusoidal fit ($\beta$ = 0.0 ± 0.01 mag/deg) and we obtain the same value with our multi-component fit. This flat slope is demonstrated by Figure 5, which shows that the residual magnitude to the multi-component fit has no significant dependence on $\theta$. While similarly flat phase curves for $\theta$ in the range 0.1 to 1.0 deg have been observed for icy satellites, the dwarf planets in the Kuiper Belt, and for the Haumea family members, most of the smaller and redder objects in the Kuiper Belt (with sizes and color similar to Manwë and Thorondor) have steeper phase curves (Rabinowitz et al. 2008, Verbiscer et al. 2019). The relatively flat phase curves for both Manwë and Thorondor require surfaces for both bodies with composition and/or scattering properties different from the general population (e.g. much higher albedo and larger ice fraction). Albedos much higher than we estimate from our multi-component model (see Table 6) would require smaller sizes and higher densities than we have estimated for both bodies, inconsistent with the shape model we have assumed. At very small phase angles ($\theta < 0.1$ deg, near the lower $\theta$ range of our observations), the icy satellites exhibit strong opposition surges (Verbiscer, French, & McGhee 2004, Buratti, Bauer, & Hicks 2011). Our observations do not extend to sufficiently low phase angle to rule out this possibility for Manwë or Thorondor.

We note here that our analysis has not allowed for independent values of $\beta$ for Manwë and Thorondor. It is possible that Thorondor has a steeper solar phase curve than the value we derive for Manwë and Thorondor together. The phase curve is sensitive to the rotational light curve we assume for Thorondor which is not well constrained by the observations (see further discussion below).

### 4.3 The Shape, Mass, Density, Size, and Albedo of Manwë

We have shown that Manwë-Thorondor has a clear periodic component with a fixed period over the entire span of the observations. To 1st order (~10% precision) and over a time span of a few days, the variability is sinusoidal. However, we obtain a better fit at all epochs (including fits to the separate Manwë and Thorondor photometry from HST) with a multi-component model where Manwë is a barbell-shaped, strengthless contact binary rotating at the dominant period (with two peaks per rotation) while Thorondor varies sinusoidally with a single-peaked period of ~309 days and amplitude ~0.55 mag. It is remarkable that we obtain this fit by tuning only 6 parameters (Thorondor's rotational period and amplitude, Manwë's spin orientation, and the rotational phases of Manwë and Thorondor). We must also make a small adjustment to the orbital period of Thorondor so that no eclipses occur, but no fine tuning is required for that purpose. It follows that our shape model is at least consistent with the observations.

If the shape model we have used is correct and Manwë is a strengthless body in hydrostatic equilibrium, then Manwë's density ($\rho \sim 0.8$ g/cm$^3$) is uniquely determined by its rotation period ($P_M \sim 11.88$ h). The equivalent spherical diameters for Manwë's volume and area ($d_{MVOL}$ = 150 km and $d_M$ = 188 km) and its mass ($m_M = 1.41 \times 10^{18}$ kg) follow directly from the known system mass ($m_{sys} = 1.941 \times 10^{18}$ kg) as long we have a good estimate for Thorondor's relative mass, $m_T/m_M$. As we discuss below, our model fit to the photometry suggests $m_T/m_M = 0.38$ (or less if Thorondor's shape is aspherical). It then follows that our estimates for $d_{MVOL}$ and $d_M$ are accurate to ~15% precision. Given the absolute magnitude we measure at maximum brightness in the F606W band, $H_M = 6.94$, and assuming Manwë's V-band magnitude is fainter than the F606W-band magnitude by $\Delta_{V,F606W} = 0.383$ mag (derived using *synphot* and Manwë's F606W-F814W color, see Benecchi et al. 2009), we can then derive an albedo for Manwë, $p_M = 0.06$. This is within the nominal range of albedos known for KBOs the size of Manwë (Brown and Butler 2017, Stansberry et al. 2008).

### 4.4 The Size, Shape, Albedo, and Relative Mass of Thorondor

Regarding the shape and rotation of Thorondor, we have no good constraints except that the brightness variability must be at least as large as we have measured (> ~0.55 mag) and the time scale of the variability must be ~300 days if it is periodic. However, we do not have enough observations to verify periodicity. Our model fit does require that the surface-area equivalent spherical diameters of Manwë and Thorondor at maximum brightness have a ratio close to the values we assume, $d_T/d_M = 108/188 = 0.57$ (assuming matching albedos). If Thorondor is spherical and has the same density as Manwë, then we derive the relative mass, $m_T/m_M = [d_T/d_{MVOL}]^3 = 0.38$. Here we use the relation $d_{MVOL}/d_M = 0.797$, which derives from the shape model we have assumed for Manwë. This estimate for $m_T/m_M$ is an upper limit if Thorondor's shape is non-spherical because a sphere has the smallest surface area to volume ratio of any shape.

From Sec 4.3, our assumption that Manwë is strengthless and in hydrostatic equilibrium allows us to determine $d_M$ with better than 15% certainty. We can then accept our estimate for Thorondor's diameter, $d_T$ = 108 km, as similarly accurate for a sphere with an albedo at maximum brightness matching Manwë's albedo. However, from the absolute magnitude we measure for Thorondor ($H_T$ = 7.73 mag), and assuming the same F606W- to V-band transformation we use for Manwë ($\Delta_{V, F606W}$ = 0.383 mag), we derive a maximum-brightness albedo $p_T$ = 0.09, significantly larger than the corresponding albedo we derive for Manwë ($p_M$ = 0.06)  This violates an implicit assumption of our model. It is not possible for both our model and our observations to be correct in this case.

The likely source of this conflict is our assumption that Thorondor's light curve is strictly periodic and sinusoidal over the entire time spanned by the observations. If we drop this assumption, we can artificially construct a light curve for Thorondor that yields a lower value for $p_T$. For example, if we alter our model to predict a brightness for Thorondor that is ~50% higher during the time period spanned by the HST observations (2007-2013) compared to later observations (2016-2018), then $H_T$ increases by 0.44 mag and $p_T$ matches the value we derive for Manwë.

This brightness change would occur naturally if Thorondor is lenticular shaped (similar to the "Ultima" lobe of 2014 MU69; Stern et al 2019), rotating about its shortest axis (i.e. the lowest-energy rotation state) and Thorondor's rotation pole is closely aligned with its orbital pole.  The orbital pole's angle, $\alpha_T$, with respect to Earth's line of site varies from 81.6 to 86.3 deg during the HST epochs, and from 90.1 to 91.4 deg during the later observed epochs. To calculate the resulting change in brightness, assume Thorondor is a triaxial ellipsoid with semi-major axes, a > b > c. With a/c = 7.33, and b/c = 6.67, we have a flattened shape similar to the Ultima lobe, but with roughly half the relative thickness. Then from Eq. (4) of Connelly and Ostro (1984), Thorondor's projected area, A, viewed from Earth is:

$$A= \pi ab[(\sin \alpha_T \sin \omega /[a/c])^2 + (\cos \alpha_T \cos \omega /[b/c])^2 + (\cos \alpha_T)^2]^{1/2},$$

where $\omega$ is the rotational phase. This yields a mean visible surface area ~40% higher during the HST epochs than the later observed epochs, nearly matching the long-term variability we require for Thorondor.

Although a lenticular shape would explain the long term changes in Thorondor's brightness, it would not explain the ~300-day variability we observe with amplitude ~0.55 mag. However, such variability might arise from the precession of the orbital pole for this shape.  From Eq. (12) of Dobrovolskis (1995), the instantaneous rotational acceleration, $d^2\omega/dt^2$, for a triaxial ellipsoid on an eccentric orbit about a spherical primary at periapsis is:

$$d^2\omega/dt^2 = -K(1-e)^6 - 6\pi\gamma T^{-2}(1-e)^3\sin(2\nu)$$

where T is the satellite's orbital period, $\gamma = (a^2-b^2)/(a^2+b^2)$, $\nu$ is the angle between the long axis and the direction to the primary, and K ~$1.6\times10^{-11}$ $yr^{-2}$ for an icy body the size of Thorondor (scaled from the value estimated by Dobrovolskis  for Nereid). The extrema for $d^2\omega/dt^2$ occurs when $\nu$ is an odd multiple of $\pi/4$. With T = 110.176 d (see Table 1) and $\gamma = 0.095$ for the lenticular shape we have assumed for Thorondor, the extrema equate to $\pm 0.0056$ $d^{-2}$. Given Thorondor's slow rotation, this torque would be enough to alter Thorondor's rotational velocity by as much as 25% in one orbit. If Thorondor's rotational pole is tilted ~12.5 deg from its orbital pole (see discussion above), then the maximum torque would precess the rotational pole by ~70 deg about the orbital pole. Although the likely outcome over many orbits would be chaotic changes in Thorondor's rotation period and pole orientation, similar to the chaotic rotation initially suggested for Pluto's smaller moons (Showalter & Hamilton 2015), the resulting oscillations in $\alpha_T$ might yield the ~300-day periodicity we observe for Thorondor at recent epochs.

### 4.5 Mutual Events

Our multi-component model does a good job of fitting the observed light curves on the assumption that no mutual events occurred during the observation windows. This is a plausible assumption given the large uncertainty in the event mid-times. However, we have explored alternative model fits for which significant mutual events do occur and significantly alter the model light curves. With fine tuning of the orbital parameters of Thorondor at values outside their ranges of uncertainty, it is possible to produce mutual event light curves that fit the observations approximately. If these solutions are correct, they would require significant secular changes to Thorondor's orbit over the time spanned by the observations. Additional HST imaging would likely be required to verify these orbital changes.

There is the additional possibility that mutual events occur during the observation windows, but the resulting alteration of the light curves is marginal. For example, if Thorondor were to eclipse or occult the narrow connecting bridge between the lobes of Manwë, this would significantly reduce the expected depth and duration of the event. This scenario might be possible with an appropriate choice for the direction of Manwë's pole (while keeping the same angle with Earth's direction). If Thorondor has a lenticular shape that we view edge on during the mutual events (see discussion, above), then this would further shorten the events and reduce their depths. It is conceivable that such an event is the cause of the dip observed in the light curve with Gemini on 2017 Jul 22 at $\omega = 0.25$ (see discussion in Sec 3.5). We plan to explore such scenarios in a subsequent paper.

Regarding the detectability of future mutual events, there is a possibility that the event season is either longer or shorter than nominally predicted. Grundy et al. (2014) assumed a spherical shape for Manwë. However, if the barbell shape we obtain for Manwë is correct, then the orientation of Manwë's rotational pole is a significant factor affecting both the duration of the events and the range of dates when the relative position of the Earth and Sun allow eclipses and occultations. Relative to the nominal diameter for the eclipse predictions (160 km) assumed by Grundy et al., Manwë may be larger by a factor of 2 in one orientation, but smaller by a factor of 2 in the perpendicular direction. The shape and orientation of Thorondor also affect the predictions. For favorable orientations, the event season may be a year or two longer than previously predicted.

## 5. **CONCLUSIONS**

We present multi-epoch light curves of the Manwë-Thorondor eclipsing system that show complex variability. Although we collected this data with the hope that we would observe a significant mutual event, we do not see clear evidence of such events at any of the epochs covered by our observations. Instead, we find a system with a light curve that is clearly periodic from day to day, but changes shape, depth and amplitude over the course of a year. The simplest explanation for not detecting the mutual events is that they did not occur when we were looking. Given their timing uncertainty, this explanation has a significant likelihood. This missed opportunity, however, has a silver lining. The absence of the mutual events allows us to construct a relatively simple, multi-component model to fit the light curve we observe. The success of our model fit suggests that Manwë may be a tidally distorted bi-lobed body with the barbell shape expected for a strengthless body in hydrostatic equilibrium. From the constraints of the shape model and from the system mass and photometric constraints, we determine Manwë's size, albedo, and density. The resulting values for these properties are typical of other KBOs and contact binaries of comparable size.

We do not have tight constraints on Thorondor's shape or size, but we can conclude that at recent epochs Thorondor was ~33% as bright as Manwë (at peak for both bodies), but had a significantly larger or brighter visible surface area during the time period covered by the earlier HST observations. This could be explained by an unusually flattened shape for Thorondor, viewed nearly edge on during the later epochs when Earth passes through Thorondor's orbit plane and the mutual events are expected. If confirmed, this shape and rotational orientation provide a valuable constraint in future efforts to determine the dynamical process that created the multi-body system.

In a subsequent paper, we hope to better explore the uncertainties of our shape model. There are likely a range of models that would work as well or better to fit the observations. Based on the range of values for the physical properties determined by these models, we could then determine their uncertainty ranges. If models are found that yield much higher albedos for Manwë and Thorondor, these would be more consistent with the unusually flat solar phase curve we determine for both bodies.

Choosing the best shape model, we would be best prepared to explore the possibility that the small deviations we observe in the residual signal are associated with marginally-occurring mutual events or an unusual shape for Thorondor. The Manwë-Thorondor system is complex, but promises to be a rich system for further exploration. Future ground- and resolved space-based observations would provide tighter constraints on the separate rotation states of Manwë and Thorondor, and may reveal properties of the system overlooked by our preliminary model.

**Acknowledgements**

We give special thanks to the telescope operators and support staff at the SOAR and Gemini telescopes who took the queued observations or helped to make remote observing possible. This work was funded in part by the National Aeronautics and Space Administration under Grant/Contract/Agreement No. NNX15AE04G issued through the SSO Planetary Astronomy Program. Parts of this research is based on data obtained at the Lowell Observatory's Discovery Channel Telescope (DCT). Lowell is a private, nonprofit institution dedicated to astrophysical research and public appreciation of astronomy and operates the DCT in partnership with Boston University, the University of Maryland, the University of Toledo, Northern Arizona University, and Yale University. Partial support of the DCT was provided by Discovery Communications. LMI was built by Lowell Observatory using funds from the National Science Foundation (grant No. AST-1005313).